\documentclass[a4paper,11pt]{article}
\pdfoutput=1
\usepackage{jheppub} % for details on the use of the package, please
\usepackage{graphicx,color,rotating}
\usepackage{hyperref}
\usepackage{epsfig,color}
\usepackage{slashed}
\usepackage{amsfonts}

\newcommand{\lsim}{\raisebox{-0.13cm}{~\shortstack{$<$ \\[-0.07cm] $\sim$}}~}
\newcommand{\gsim}{\raisebox{-0.13cm}{~\shortstack{$>$ \\[-0.07cm] $\sim$}}~}

\begin{document}
\title{Heating neutron stars with GeV dark matter}

\renewcommand{\thefootnote}{\arabic{footnote}}

\author{
Wai-Yee Keung$^{1}$,  Danny Marfatia$^{2}$, and
Po-Yan Tseng$^{3}$}
\affiliation{
$^1$ Department of Physics, University of Illinois at Chicago,
Illinois 60607 USA \\
%%
%$^2$ Physics Division, National Center for Theoretical Sciences,
%Hsinchu, Taiwan \\
%%
$^2$ Department of Physics and Astronomy, University of Hawaii,
Honolulu, HI 96822, USA \\
$^3$ Department of Physics and IPAP, Yonsei University,
Seoul 03722, Republic of Korea \\
}
%\pacs{14.80.Bn.,14.80.Da,14.80.Ec}
\date{\today}

\abstract{
An old neutron star (NS) may capture halo dark matter (DM) and get heated up 
by the deposited kinetic energy, thus behaving like a thermal DM detector with
sensitivity to a wide range of DM masses and a variety of
DM-quark interactions. Near future infrared telescopes will measure NS temperatures
down to a few thousand Kelvin and probe NS heating by DM capture.
We focus on GeV-mass Dirac fermion DM (which is beyond the reach of current 
DM direct detection experiments) in scenarios in which the DM capture
rate can saturate the geometric limit. For concreteness, we study
(1) a model that invokes dark decays of the neutron to explain the neutron lifetime anomaly, and
(2) a framework of DM coupled to quarks through a vector current portal.
In the neutron dark decay model a NS can have a substantial DM population, 
so that the DM capture rate can reach the geometric limit 
through DM self-interactions even if the DM-neutron scattering cross section is tiny. 
We find NS heating to have greater sensitivity than 
multi-pion signatures in large underground detectors for the neutron dark decay model,
and sub-GeV gamma-ray signatures for the quark vector portal model.
}

\maketitle
%\newpage
\section{Introduction}

Dark matter (DM)  may have a variety of interactions
with SM particles and with DM itself, but with strengths
that have evaded observation.
A neutron star (NS) orbits through large fluxes of halo DM particles 
which may lose their energy via their interactions with the NS 
and become gravitationally bound to it. 
The high density and strong gravity of a NS
may be able to compensate the feeble DM interactions 
and  enhance the DM capture rate. The capture of halo DM by a NS had been extensively studied~\cite{Guver:2012ba,Chen:2018ohx,McDermott:2011jp,Garani:2018kkd,Bell:2019pyc,Garani:2019fpa,Baryakhtar:2017dbj}. 
During the capture process, the strong gravitational potential of the NS accelerates 
the DM to more than half the speed of light, and DM-neutron 
scattering releases this kinetic energy to heat up the NS. 
Consequently, the NS temperature evolution will deviate 
from the standard cooling profile.
A possible observable signal of DM capture by a NS is the detection of unexpectedly hot old neutron stars.
The temperature of an old neutron star can be heated by
$\sim 100$~K to $\sim 2000$~K,
which is within the near-infrared band of the blackbody spectrum.
The thermal emissions from nearby (within 100~pc), faint and isolated NS 
can be probed by upcoming infrared telescopes such as the James Webb 
Space Telescope (JWST), the Thirty Meter Telescope, and the European Extremely 
Large Telescope~\cite{Baryakhtar:2017dbj}.
 
A DM-neutron cross section of $\sim 2\times 10^{-45}~{\rm cm^2}$ 
is large enough to heat up an old neutron star to
$\sim 1000$~K for DM masses between GeV and PeV. 
For DM lighter than a GeV, the capture rate is suppressed by
Pauli blocking, while for DM heavier than a PeV, multiple scattering 
is necessary to slow down the halo DM particles. However, the total capture rate 
must lie below the geometric limit, which corresponds to all the ambient halo DM within
the geometric area of the NS being captured.

We study scenarios with three aspects: (1) the DM is of GeV mass, which makes direct detection problematic,
(2) the DM is a Dirac fermion, so that it matters whether the particle or the antiparticle is the DM, and (3) the DM
capture rate can reach the geometric limit. 
Specifically, we  examine NS heating in the neutron dark decay model~\cite{Fornal:2018eol,Grinstein:2018ptl} and in a
quark vector current portal framework~\cite{Berger:2019aox,Kumar:2018heq}. 

The neutron dark decay model finds its origin in the recent neutron lifetime anomaly which is a $\sim 4\sigma$ discrepancy~\cite{Patrignani:2016xqp} in the 
neutron lifetimes measured in beam~\cite{Byrne:1996zz,Yue:2013qrc} 
and bottle~\cite{Pichlmaier:2010zz,Steyerl:2012zz,Arzumanov:2015tea} experiments.
If the neutron has the dark decay, $n \to \chi+\phi$, where $\chi$ and $\phi$ are dark sector particles, with a partial width of about 
$7.1\times 10^{-30}$~GeV  the discrepancy is alleviated. The scalar $\phi$ is almost massless and no heavier than an MeV. The DM particle is very slightly lighter than the neutron and is a Dirac fermion to avoid constraints from neutron-antineutron oscillations. 
Multi-pion signatures in neutron-antineutron oscillation searches by 
Super-Kamiokande only constrain
the model if the DM is $\bar\chi$~\cite{Keung:2019wpw}. 
The model is interesting in that, as we will see,
a NS can be composed of a substantial DM population, 
so that the DM capture rate can reach the geometric limit 
through DM self-interactions even if the 
DM-neutron scattering cross section is small.
%The typical DM-neutron cross section is about $10^{-49}~{\rm cm^2}$, which is too weak to heat a NS 
%to an observable level. However, DM self-capture helps to enhance the capture rate due to substantial DM converted from neutron inside NS.
% After solving the equation of state (EoS), when $m_\phi \gtrsim 0.2$ eV, NS is at the mixed phase, and $\chi$ can contribute about 40\% of NS.
%Therefore, the DM-self capture can dominates the capture rate  due to that the much larger DM-self scattering cross section of 
%$\mathcal{O}(10^{-40})~{\rm cm^2}$.  Therefore, the DM-self interaction from the box diagram is crucial 
%to make capture rate exceed geometric limit and heat up old NS temperature more than $\mathcal{O}(1500)$ K.
%On the other hand, for $m_\phi \lesssim 0.2$ eV, NS is at neutron phase, such that DM captured by neutron is the main process, and NS 
%temperature only increase to $\simeq 400$ K. 

As a second example, we consider dark matter that couples to $u,d,s$ quarks through a dimensional-6 vector portal 
with independent couplings $\alpha_{u,d,s}$. These couplings can be chosen so that the DM capture rate reaches the geometric limit.
The NS also gets heated by the annihilation of GeV DM to light mesons (which can be  described by chiral perturbation theory~\cite{Berger:2019aox,Kumar:2018heq}).

%For the effective couplings to first generation quarks of 
%$\alpha_{u,d}/\Lambda \simeq 10^{-6}~{\rm GeV^{-1}}$, 
%old NS temperature is heated up to $\simeq 2500$ K.
%Which is more stringent than the projected sensitivity 
%from DM indirect detection using the sub-GeV gamma-ray 
%$\alpha_{u,d}/\Lambda \simeq 10^{-2}~{\rm GeV^{-1}}$.
%

The paper is organized as follows. 
In section~\ref{sec:ns_capture}, we review the process of
DM capture by a NS,
and the resultant NS temperature evolution is described in section~\ref{sec:ns_evol}.
We study the neutron dark decay model in section~\ref{sec:neutron_dark},
and the quark vector current portal model in section~\ref{sec:vector_portal}.
We summarize our results in section~\ref{sec:summary}.

\bigskip

\section{Dark matter capture in neutron stars}
\label{sec:ns_capture}

DM capture by a NS is primarily governed by DM-nucleon scattering
and by DM self-interactions if a significant DM population is bound by the NS. 
For weak scale DM, there are stringent upper limits on the DM-nucleon cross section, 
but constraints on DM self-interactions are relatively loose.
Interestingly, the preferred range for the self-interaction cross section to alleviate the core-cusp problem
is $0.1~{\rm cm^2/g}\lesssim \sigma_{\chi \chi}/m_\chi\lesssim 1~{\rm cm^2/g}$~\cite{Tulin:2012wi}.
This corresponds to $\sigma_{\chi \chi}\simeq 10^{-24} {m_\chi \over {1~\rm{GeV}}}~{\rm cm^2}$,
which is much weaker than the upper limit 
$\sigma_{\chi-{\rm nucleon}}\lesssim 10^{-38}~{\rm cm^2}$ from DM direct detection experiments~\cite{Abdelhameed:2019hmk}.
Therefore, DM self-interactions may dramatically enhance the capture rate.
Other processes, like DM-neutron annihilation, $\chi\bar\chi$ annihilation
and neutron decays to DM, also affect DM capture, and
are included in our discussion below which is tailored for the neutron dark decay model; the corresponding 
equations for the quark vector current portal scenario are simpler and obtainable by straightforward modifications.

Because we study scenarios of Dirac fermion DM, the DM particle is either $\chi$ or $\bar{\chi}$.
We consider the general case in which the NS is composed of both neutrons and $\chi$,
as is the case for the neutron dark decay model we consider.
The evolution of the number of DM particles $N_{\rm DM}$ in the neutron star  is described by~\cite{Zentner:2009is}
\begin{eqnarray}
\label{eq:DMnumber}
\frac{dN_{\rm DM}}{dt}=\left\lbrace
\begin{array}{l}
C_c+C^{\chi \chi}_s(N_{\rm DM}+N_\chi)\,, ~~~~~~~~~~~~~~~~~~~~~~~~~~~~~~~~~~~~~~~~~~~~~~\text{if DM is $\chi$} \\
C_c+(C^{\bar{\chi} \bar{\chi}}_s N_{\rm DM}+ C^{\bar{\chi} \chi}_s N_\chi)-C^{\bar{\chi}n}_a N_{\rm DM}N_n-C_a N_{\rm DM} N_\chi\,, ~~~~\text{if DM is $\bar{\chi}$}
\end{array}
\right.
\end{eqnarray}
where we distinguish the component $N_\chi$ produced by neutron decay,  $n \to \chi+\phi$, from the
halo DM component $N_{\rm DM}$ because they may have different thermal properties.
We assume that the rate of $n \to \chi+\phi$ 
is large enough to keep the neutrons and $\chi$ in thermal equilibrium.
Halo DM-neutron elastic scattering contributes to the capture rate, and if DM is $\bar{\chi}$, halo DM also annihilates with neutrons, which under the assumption of a uniform mass distribution, are respectively given by~\cite{McDermott:2011jp}
\begin{eqnarray}
\label{eq:Cc}
C_c &=& \sqrt{\frac{6}{\pi}}\frac{\rho_{\rm DM}}{m_\chi}\frac{v^2_{\rm esc}(R)}{\bar{v}^2}
(\bar{v}\xi \sigma^{\rm elastic}_{{\rm DM-}n})
 N_n 
\left(1-\frac{1-e^{-B^2}}{B^2} \right)\,, \nonumber \\
C^{\rm ann} &=& \sqrt{\frac{6}{\pi}}\frac{\rho_{\rm DM}}{m_\chi}\frac{v^2_{\rm esc}(R)}{\bar{v}^2}
(\bar{v}\sigma^{\rm ann}_{\bar{\chi} n})
 N_n 
\left(1-\frac{1-e^{-B^2}}{B^2} \right)\,,
\end{eqnarray}
where the escape velocity of the NS is
$v_{\rm esc}(R)=\sqrt{2GM/R}\simeq 0.63\,c$,
 $\bar{v}$ is the DM dispersion velocity, and
$\rho_{\rm DM}$ is the local DM density;  the relevant parameter values for the NS and the DM halo are listed in the table below.
$N_n$ is the total number of neutrons in the NS, and $B^2\equiv(3/2)(v^2_{\rm esc}/\bar{v}^2)\beta_-$ with 
$\beta_-=4m_\chi m_n/(m_\chi-m_n)^2$
appears after averaging over the DM velocity distribution. Of course, $m_{\rm DM} \equiv m_\chi=m_{\bar\chi}$.

\begin{center}
\begin{tabular}{ll}
%\hline  \hline
% &  Value   \\
  \hline \hline
%  Gravitational constant \hspace{0.2in} & $G=6.674\times 10^{-8}~{\rm cm^3 g^{-1} s^{-2}}$ \\
% Light speed \hspace{0.2in} & $c=2.9979\times 10^{10}~{\rm cm/s}$ \\
%  Boltzmann constant \hspace{0.2in} & $k_B=8.617\times 10^{-14}~{\rm GeV\,K^{-1}}$ \\
%  Planck constant \hspace{0.2in} & $h/2\pi=6.582\times 10^{-25}~{\rm GeV\cdot s}$ \\
%  Neutron mass \hspace{0.2in} & $m_n=1.6749\times 10^{-24}~{\rm g}=0.9396~{\rm GeV}$ \\
%  Proton mass \hspace{0.2in} & $m_p=1.6726\times 10^{-24}~{\rm g}=0.9383~{\rm GeV}$ \\
% Electron mass \hspace{0.2in} & $m_e=9.1094\times 10^{-28}~{\rm g}=0.511\times 10^{-3}~{\rm GeV}$ \\
%  Solar mass \hspace{0.2in} & $M_\odot=1.98892\times 10^{33}~{\rm g}$ \\
  Velocity dispersion of DM \hspace{0.2in} & $\bar{v}=270~{\rm km/s}$ \\
  Local DM density \hspace{0.2in} & $\rho_{\rm DM}=0.4~{\rm GeV/cm^3}$ \\
  NS  velocity relative to GC \hspace{0.2in} & $v_N=220~{\rm km/s}$ \\
  NS  mass \hspace{0.2in} & $M=1.44 M_\odot=2.86\times 10^{33}~{\rm g}$ \\
  NS  radius \hspace{0.2in} & $R=10.6~{\rm km}$ \\
  NS  fermion density \hspace{0.2in} & $\rho_F=5.7\times 10^{14}~{\rm g/cm^3}$ \\
  NS  fermion number density \hspace{0.2in} & $n_F=3.4\times 10^{38}~{\rm cm^{-3}}=2.125\, n_0$ \\
  \hline
   \hline
\end{tabular}
\end{center}

\bigskip

We assume that the neutrons inside the NS behave as a Fermi gas and estimate the Fermi momentum to be
$p_F\simeq(3\pi^2\rho_F/m_n)^{1/3}=437$~MeV.
DM-neutron scattering only occurs when the momentum exchange $\delta p$
is larger than $p_F$.
We take this Pauli blocking into account by introducing a factor $\xi={\rm min}(\delta p/p_F,1)$
in the above capture rate $C_s$.
%
% in the effective DM-neutron cross section 
%$\bar{\sigma}^{\rm elastic}_{\rm{DM-} n}\equiv{\rm min}(\xi\sigma^{\rm elastic}_{\rm{DM-}n},\sigma_{\rm crit}\,r_\sigma)$
%and the effective annihilation cross section
%$\bar{\sigma}^{\rm ann}_{\bar{\chi} n}\equiv 
%{\rm min}\left(\sigma^{\rm ann}_{\bar{\chi} n},\sigma_{\rm crit}\,(1-r_\sigma)\right)$, where $r_\sigma\equiv \xi \sigma^{\rm elastic}_{\rm{DM-} n}
%/(\xi\sigma^{\rm elastic}_{\rm{DM-}n}+\sigma^{\rm ann}_{\bar{\chi} n})$
%gives the fractional contribution from DM-neutron elastic scattering.
%
Note that once the sum of cross sections 
($\xi \sigma^{\rm elastic}_{\chi n}$ for $\chi$ DM 
, or
$\xi \sigma^{\rm elastic}_{\bar{\chi}n}+\sigma^{\rm ann}_{\bar{\chi}n}$ 
for $\bar{\chi}$ DM)
is larger than
critical cross section, $\sigma_{\rm crit}=\pi R^2m_n/M$, and the sum of 
the capture rate and annihilation rates cannot be larger than
the geometric limit, i.e., $C_c+C^{\rm ann} \leq C_c|_{\rm geom}$.
This is equivalent to $N_n (\xi \sigma^{\rm elastic}_{\chi n})\leq \pi R^2$ if DM is $\chi$, and
$N_n (\xi \sigma^{\rm elastic}_{\bar{\chi}n}+\sigma^{\rm ann}_{\bar{\chi}n})\leq \pi R^2$ if DM is $\bar{\chi}$.
For 1~GeV $\chi$ DM, the geometric limit 
$C_c|_{\rm geom}\simeq 8.2\times 10^{32}\,{\rm yr^{-1}}$
corresponds to $\sigma_{\rm crit}\simeq 10^{-45}~{\rm cm^2}$~\cite{Garani:2018kkd}.

The DM capture rate due to scattering
 on $\chi$ from neutron conversion inside the NS  or on the trapped DM (whose population is negligible in comparison) is~\cite{Guver:2012ba} 
\begin{eqnarray}
C^{\chi\chi}_s=C^{\bar{\chi}\bar{\chi}}_s &=&
\sqrt{\frac{3}{2}}\frac{\rho_{\rm DM}}{m_\chi}
\sigma_{\chi\chi \to \chi \chi}v_{\rm esc}(R)\frac{v_{\rm esc}(R)}{\bar{v}}
%\left\langle \hat{\phi_\chi}\right\rangle 
\frac{{\rm erf}(\eta)}{\eta}\frac{1}{1-\frac{2GM}{R}}\,, \nonumber \\
C^{\bar{\chi}\chi}_s &=& \sqrt{\frac{3}{2}}\frac{\rho_{\rm DM}}{m_\chi}
\sigma_{\bar{\chi}\chi \to \bar{\chi} \chi}v_{\rm esc}(R)\frac{v_{\rm esc}(R)}{\bar{v}}
%\left\langle \hat{\phi_\chi}\right\rangle 
\frac{{\rm erf}(\eta)}{\eta}\frac{1}{1-\frac{2GM}{R}}\,,
\end{eqnarray}
where we have again assumed that the mass density of the NS is uniform.
%with conservative value of $\left\langle \hat{\phi_\chi}\right\rangle=1$.
Here, $\eta=\sqrt{3/2}(v_N/\bar{v})$, 
with $v_N$ the NS velocity relative to the Galactic center.
For these cases, we define the geometric limits,
$N_\chi \sigma_{\chi\chi \to \chi \chi}\leq
\pi R^2$ 
and $N_\chi \sigma_{\bar{\chi}\chi \to \bar{\chi} \chi}\leq
\pi R^2$.
The trapped DM with velocity $v_{\rm DM}$ will form its own sphere of radius $r_{\rm DM}(t)$,
and the evolution of $r_{\rm DM}(t)$ is derived 
as follows.
The kinetic energy of each DM particle can be expressed in terms of the orbital radius $r_{\rm DM}(t)$ as~\cite{Guver:2012ba}
\begin{equation}
\label{eq:10}
E_{\rm DM}=\frac{2\pi}{3}G\rho_F m_\chi r^2_{\rm DM}=\frac{1}{2}m_\chi v^2_{\rm DM}\,,
\end{equation}
with the rate of change in kinetic energy given by~\cite{McDermott:2011jp} 
\begin{eqnarray}
\label{eq:11}
\frac{dE_{\rm DM}}{dt}=\left\lbrace
\begin{array}{cc}
 -\xi' 
\left[
n_F(1-a_\chi) \sigma^{\rm elastic}_{\chi n}
+n_F a_\chi \sigma_{\chi\chi \to \chi \chi}
\right]  v_{\rm DM}
\delta E\cdot 
{\rm sign}(T_{\rm DM}-T_{\rm int}) & \nonumber \\
+C^{\chi \chi}_s\Delta E\,,~~~~~~~~~~~~~~~~~~~~~~~~~~~~~~~~~~~~~~~~~~~~~~~~~~~~~~~~~~~~~~~~~~~~~~ &\text{if DM is $\chi$} \nonumber \\
-\xi' 
\left[
n_F(1-a_\chi) \sigma^{\rm elastic}_{\bar{\chi} n}
+n_F a_\chi \sigma_{\bar{\chi}\chi \to \bar{\chi} \chi}
\right]  v_{\rm DM}
\delta E\cdot 
{\rm sign}(T_{\rm DM}-T_{\rm int}) & \nonumber \\
+C^{\bar{\chi}\bar{\chi}}_s\Delta E\,,~~~~~~~~~~~~~~~~~~~~~~~~~~~~~~~~~~~~~~~~~~~~~~~~~~~~~~~~~~~~~~~~~~~~~~ &\text{if DM is $\bar{\chi}$}
\end{array}
\right.  \\
\end{eqnarray}
where $a_\chi$ is the fractional number of $\chi$ in the NS, 
and $1-a_\chi$ is the fractional number of neutrons in the NS.
The first (second) term in brackets corresponds to an energy release
$\delta E = 2m_r E_{\rm DM}/(m_n+m_\chi)$ to the neutron component ($\chi$ component) of the NS~\cite{Guver:2012ba}, where $m_r$ is the
reduced mass of the DM-neutron system.\footnote{
The analytic expression for $\delta E$ is a valid approximation only if the
DM particle is much more energetic than the neutron,
and $m_\chi \sim m_n$~\cite{Guver:2012ba}.
Equation~(\ref{eq:11}) is used to determine if the trapped DM 
and neutron can achieve thermal equilibrium, a condition that is easily satisfied in the neutron dark decay model.
Therefore, this approximation has little effect on our results.  
}
The energy gain,
$\Delta E=\frac{1}{2}m_\chi(v^2_{\rm esc}-v^2_{\rm DM})$,
results from a drop in the halo DM's potential energy from 
$\frac{1}{2}m_\chi v^2_{\rm esc}$ 
to $\frac{1}{2}m_\chi v^2_{\rm DM}$ after thermalizing with the trapped DM.
Here,
$$
\frac{1}{2}m_\chi v^2_{\rm esc}=\frac{GMm_\chi}{R}+\frac{GMm_\chi}{R^3}
\left(\frac{R^2-r^2_{\rm DM}}{2} \right)\,.
$$ 
Effects of Pauli blocking are included by the factor,
$\xi'={\min}(\sqrt{2}m_r v_{\rm DM}/p_F,1)$.
The evolution of $r_{\rm DM}(t)$ is obtained by combining Eqs.~(\ref{eq:10}) and (\ref{eq:11}), and
and the temperature of the DM sphere $T_{\rm DM}$ is given by $\frac{3}{2} kT_{\rm DM}(t)=E_{\rm DM}$.

The last two terms in the second equation in Eq.~(\ref{eq:DMnumber}) depends on the DM-neutron and DM-antiDM annihilation rates~\cite{Zentner:2009is} 
\begin{equation}
C^{\bar{\chi}n}_a \simeq \frac{\left\langle 
\sigma^{\rm ann}_{\bar\chi n} v_{\rm DM} \right\rangle}{4\pi R^3/3}
\,, \quad
C_a \simeq \frac{\left\langle 
\sigma^{\rm ann}_{\bar\chi \chi} v_{\rm DM} \right\rangle}{4\pi R^3/3}\,,
\end{equation}
which depletes the total number of trapped DM.

\section{Temperature evolution}
\label{sec:ns_evol}

Soon after a NS is formed in a supernova explosion,
its core has a temperature of about $10^{11}$~K. It then 
 cools down to $10^8$~K through neutrino emission 
in about $10^5$~years.
When the core temperature falls below $10^8$~K, 
photon emission dominates the cooling process.
Unlike neutrino cooling, whose detailed mechanism is still under debate,
photo cooling has less uncertainty,
and we focus on this period of a neutron star's life.

The interior temperature $T_{\rm int}$ of a NS evolves according to~\cite{Chen:2018ohx}
\begin{equation}
\frac{dT_{\rm int}}{dt}=\frac{-\epsilon_\nu-\epsilon_\gamma+\epsilon_{\rm DM}}{c_V}\,,
\end{equation}
where $\epsilon_{\nu,\gamma,{\rm DM}}$ are the neutrino, photon and DM emissivities, and $c_V$ is the NS heat capacity
per unit volume.
Treating neutrons and the $\chi$ from neutron conversion as ideal Fermi gases, $c_V$ is given by~\cite{Shapiro:1983,Kouvaris:2007ay}
\begin{equation}
c_V=\frac{k^2_B T_{\rm int}}{3}\sum_{i=\chi,n}p_{F,i}
\sqrt{m^2_i+p^2_{F,i}}\,,
\end{equation}
where the Fermi momenta are
\begin{eqnarray}
p_{F,\chi}&=&
0.34~{\rm GeV}\left(\frac{n_F a_\chi}{n_0} \right)^{1/3}
\,, \nonumber \\
p_{F,n}&=&
0.34~{\rm GeV}\left(\frac{n_F(1-a_\chi)}{n_0} \right)^{1/3}
\,.
\end{eqnarray}

The neutrino emissivity is~\cite{Kouvaris:2007ay,Shapiro:1983}
$$
\epsilon_\nu\simeq 1.81\times 10^{-27}~{\rm GeV^4yr^{-1}}
\left(\frac{n_F}{n_0}\right)^{2/3}
\left(\frac{T_{\rm int}}{10^7~{\rm K}} \right)^8\,\,,
$$
where $n_0= 0.16~{\rm fm^{-3}}=0.16\times 10^{39}~{\rm cm^3}$, and $n_F$ is the average
fermion number density in a NS.\footnote{
Since the neutron radius is $\sim 1$~fm, $n_0$ 
sets the scale for the critical density of a NS.
A NS with central density of $6n_0$ has a $\sim 2M_\odot$ mass
which depends on the nuclear equation of state.}
Since neutrino emission depends on the eighth power of $T_{\rm int}$,
neutrinos easily escape the NS when it is young.
%the inner temperature is more relevant for the neutrino cooling.
The surface temperature $T_{\rm sur}$ of a NS is related to $T_{\rm int}$ via~\cite{Page:2004fy,Gudmunsson:1982,Gudmunsson:1983}
\begin{eqnarray}
T_{\rm sur}=\left\lbrace  
\begin{array}{l} 0.87\times 10^{6}~{\rm K}
\left(\frac{g_s}{10^{14}~{\rm cm\,s^{-2}}} \right)^{1/4}
\left(\frac{T_{\rm int}}{10^8~{\rm K}} \right)^{0.55},\quad T_{\rm int}\gtrsim 3700~{\rm K}\,\\
T_{\rm int}\, ,~~~~~~~~~~~~~~~~~~~~~~~~~~~~~~~~~~~~~~~~~~~~~~~~~~
T_{\rm int}\lesssim 3700~{\rm K}
\end{array}
\right.
\end{eqnarray}
where $g_s=GM/R^2=1.85\times 10^{14}~{\rm cm\,s^{-2}}$ is the gravitational acceleration at the surface of the NS.
Including the effect of gravitational redshift, the observed temperature $T_{\rm obs}$ at infinity is~\cite{Bell:2018pkk}
$$
T_{\rm obs}=T_{\rm sur} \sqrt{1-\frac{2GM}{Rc^2}}\,.
$$
The NS luminosity $L_\gamma$ from the outer envelope is given by the Stefan-Boltzmann law:
\begin{equation}
L_\gamma=4\pi R^2 \sigma_{\rm SB}T^4_{\rm sur}
\simeq 5.00\times 10^{11}~{\rm GeV\,s^{-1}}\left(\frac{T_{\rm sur}}{\rm K} \right)^4\,,
\end{equation}
where $\sigma_{\rm SB}=3.5383\times 10^{-2}~{\rm GeV\,cm^{-2}\,s^{-1}\,K^{-4}}$ is the  Stefan-Boltzmann constant.
Then the effective photon emissivity is
\begin{eqnarray}
\epsilon_\gamma = \frac{L_\gamma}{4\pi R^3/3} \simeq \left\lbrace
\begin{array}{l}
  2.59\times 10^{-17}~{\rm GeV^4\,yr^{-1}}\left(\frac{T_{\rm int}}{10^8~{\rm K}} \right)^{2.2}\,, \quad T_{\rm int}\gtrsim 3700~{\rm K} \nonumber \\
  2.44\times 10^{-9}~{\rm GeV^4\,yr^{-1}}\left(\frac{T_{\rm int}}{10^8~{\rm K}} \right)^4\,, \quad~~~ T_{\rm int}\lesssim 3700~{\rm K}\,.
\end{array}
\right.
\end{eqnarray}
Photon emission dominates the cooling process after $10^5$ years, when $T_{\rm obs}\lesssim 10^{6}$~K.

Dark matter can inject energy into a NS in several ways.
Halo DM-neutron elastic scattering and halo DM-neutron annihilation (if the DM is $\bar\chi$) contribute energy,
\begin{eqnarray}
\mathcal{K}_{\rm DM} = \left\lbrace
\begin{array}{c}
C_c \langle E_R \rangle\,,~~~~~~~~~~~~~~~~~~~~~~~~~~~\text{if DM is $\chi$} \\
C_c \langle E_R \rangle
+C^{\rm ann}\, (m_\chi+m_n)
\,, ~~\text{if DM is $\bar{\chi}$}
\end{array}
\right. \nonumber
\end{eqnarray}
where 
$$\langle E_R \rangle \equiv 
\frac{\int^{1}_{-1}d\cos \theta_{\rm cm} \,E_R\, \frac{d\sigma_{{\rm DM}-n}}{d\cos \theta_{\rm cm}}}{\int^{1}_{-1}d\cos \theta_{\rm cm}\, \frac{d\sigma_{{\rm DM}-n}}{d\cos \theta_{\rm cm}}} 
\simeq \frac{(1-\bar{B})m_\chi \bar{\mu}}{\bar{B}+2\sqrt{\bar{B}}\bar{\mu}+\bar{B}\bar{\mu}^2}\,,$$ 
is the angular average recoil energy transferred from the DM to a neutron in a single collision~\cite{Bell:2018pkk}.
Here, $\bar{B}\equiv 1-2GM/(c^2R)\simeq 0.60$ and $\bar{\mu}\equiv m_\chi/m_n$. 
For $m_\chi\simeq m_n$ we find $\langle E_R \rangle\simeq 0.15 m_\chi$, which implies 
that annihilation is more efficient than elastic scattering at heating a NS
if the halo DM-neutron annihilation and elastic scattering rates are comparable.

Another source of heat is the annihilation of trapped DM.
If the trapped DM is $\bar\chi$, it can annihilate with $\chi$ from neutron conversion 
or with neutrons into SM particles and inject energy,
\begin{eqnarray}
\label{eq:ann_heating}
\mathcal{E}_{\rm DM}=\left\lbrace
\begin{array}{c}
0\,, ~~~~~~~~~~~~~~~~~~~~~~~~~~~~~~~~~~~~~~~~~~~~~~~~~~~~~~~~~~~~\text{if DM is $\chi$} \nonumber \\
2m_\chi C_a N_{\rm DM} N_\chi f_{\rm DM}
+(m_n+m_\chi)C^{\bar{\chi}n}_a N_{\rm DM} N_n
\,,~~\text{if DM is $\bar{\chi}$}
\end{array}
\right.
\end{eqnarray}
where $f_{\rm DM}\subset [0,1]$ is the efficiency with which energy 
is absorbed by the NS and depends on the annihilation final states. 
For instance, $f_{\rm DM}=0$ for a purely neutrino final state,
and $f_{\rm DM}=1$ for a $\gamma \gamma$ final state. In principle, the contribution from $\bar\chi$-neutron annihilation
also has an efficiency factor, but we approximate this to unity for the final states we consider later; this also applies to the annihilation
term in $\mathcal{K}_{\rm DM}$ above.

The trapped DM also releases its energy via elastic scattering with neutrons and with $\chi$ from neutron conversion:
\begin{eqnarray}
\label{eq:24}
\mathcal{F}_{\rm DM}= \left\lbrace
\begin{array}{c}
\xi' 
\left[
n_F(1-a_\chi) \sigma^{\rm elastic}_{\chi n}
+n_F a_\chi \sigma_{\chi \chi \to \chi \chi}
\right]v_{\rm DM}
 \delta E\, N_{\rm DM} \cdot 
{\rm sign}(T_{\rm DM}-T_{\rm int}) \,, ~~\text{if DM is $\chi$} 
\nonumber \\
\xi' 
\left[
n_F(1-a_\chi) \sigma^{\rm elastic}_{\bar{\chi}n}
+n_F a_\chi \sigma_{\bar{\chi} \chi \to \bar{\chi} \chi}
\right]v_{\rm DM}
 \delta E\, N_{\rm DM} \cdot 
{\rm sign}(T_{\rm DM}-T_{\rm int}) \,, ~~\text{if DM is $\bar{\chi}$}
\end{array}
\right.\,.  \\
\end{eqnarray}
From Eqs.~(\ref{eq:11}) and (\ref{eq:24}), we see the path of energy conduction. 
The kinetic energy lost by halo DM to become trapped is transferred to the NS through scattering processes.

Summing over the above three contributions, the total DM emissivity is
\begin{equation}
\epsilon_{\rm DM}=
\frac{\mathcal{K}_{\rm DM}+\mathcal{E}_{\rm DM}+\mathcal{F}_{\rm DM}}{4\pi R^3/3}\,.
\end{equation}
%
%We conservatively set $a_\chi=0$ to obtain $c_V$.

\begin{figure}[t!]
\centering
\includegraphics[height=4.2in,angle=270]{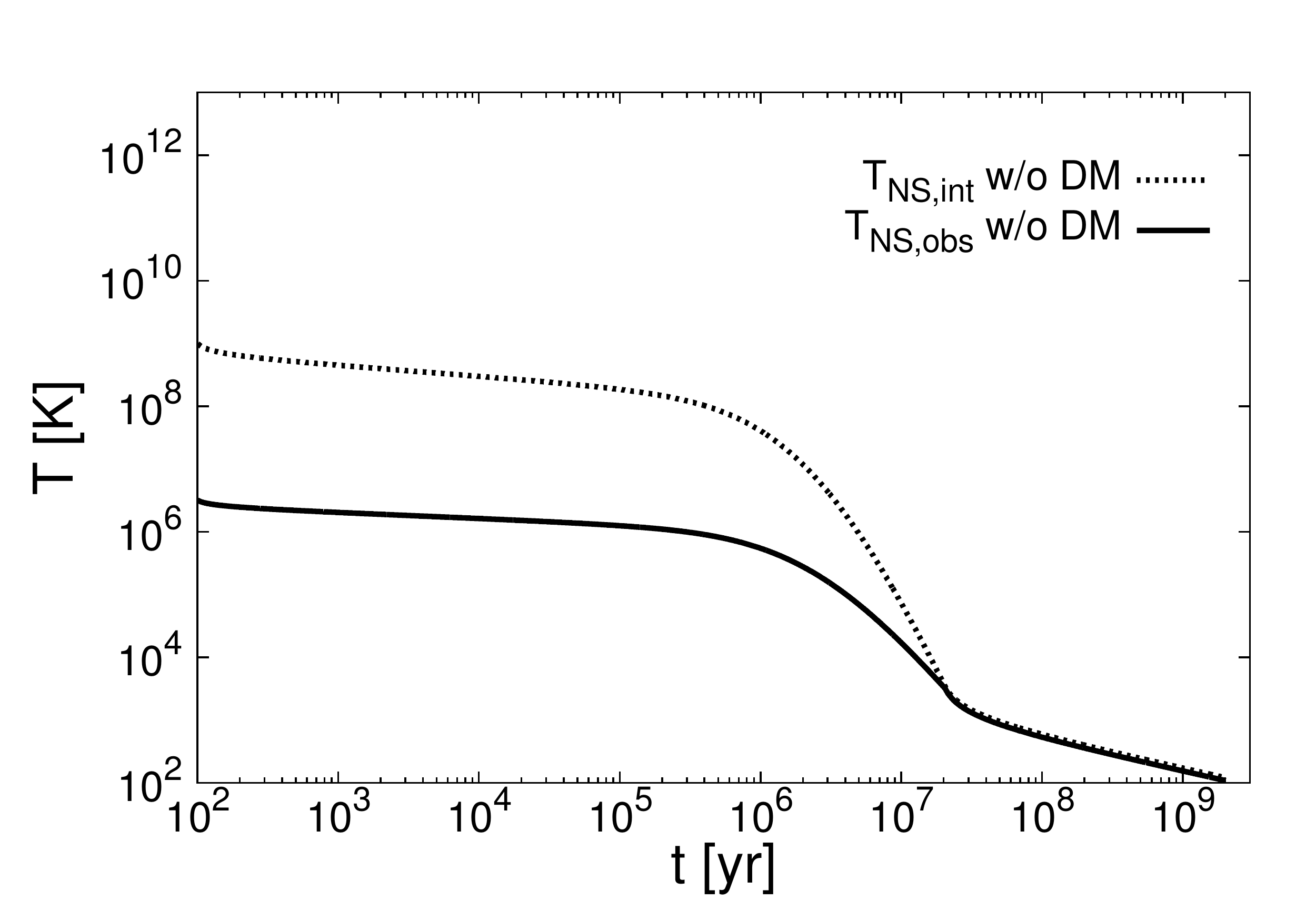}
\caption{\small \label{fig:T_NS}
The time evolution of the interior and observed NS temperatures without DM capture.
}
\end{figure}

The time evolution of the interior and observed temperatures of a NS without 
DM heating are shown in Fig.~\ref{fig:T_NS}.
For an old NS of age between $10^{8}$ and $10^{9}$~years, 
the temperature falls to about $500$~K and $150$~K, respectively.

In the rest of this section we do not consider the possibility of neutron conversion to $\chi$ and 
DM-neutron annihilation. 
Neutron star heating by DM capture can compensate the cooling
from photon emission  once  $T_{\rm int}$ falls to $\sim1000$~K.
The NS can be heated by two processes: 
$i$) kinetic heating by the captured DM, and $ii$) DM annihilation into SM final states.

\begin{figure}[t!]
\centering
\includegraphics[height=2.9in,angle=270]{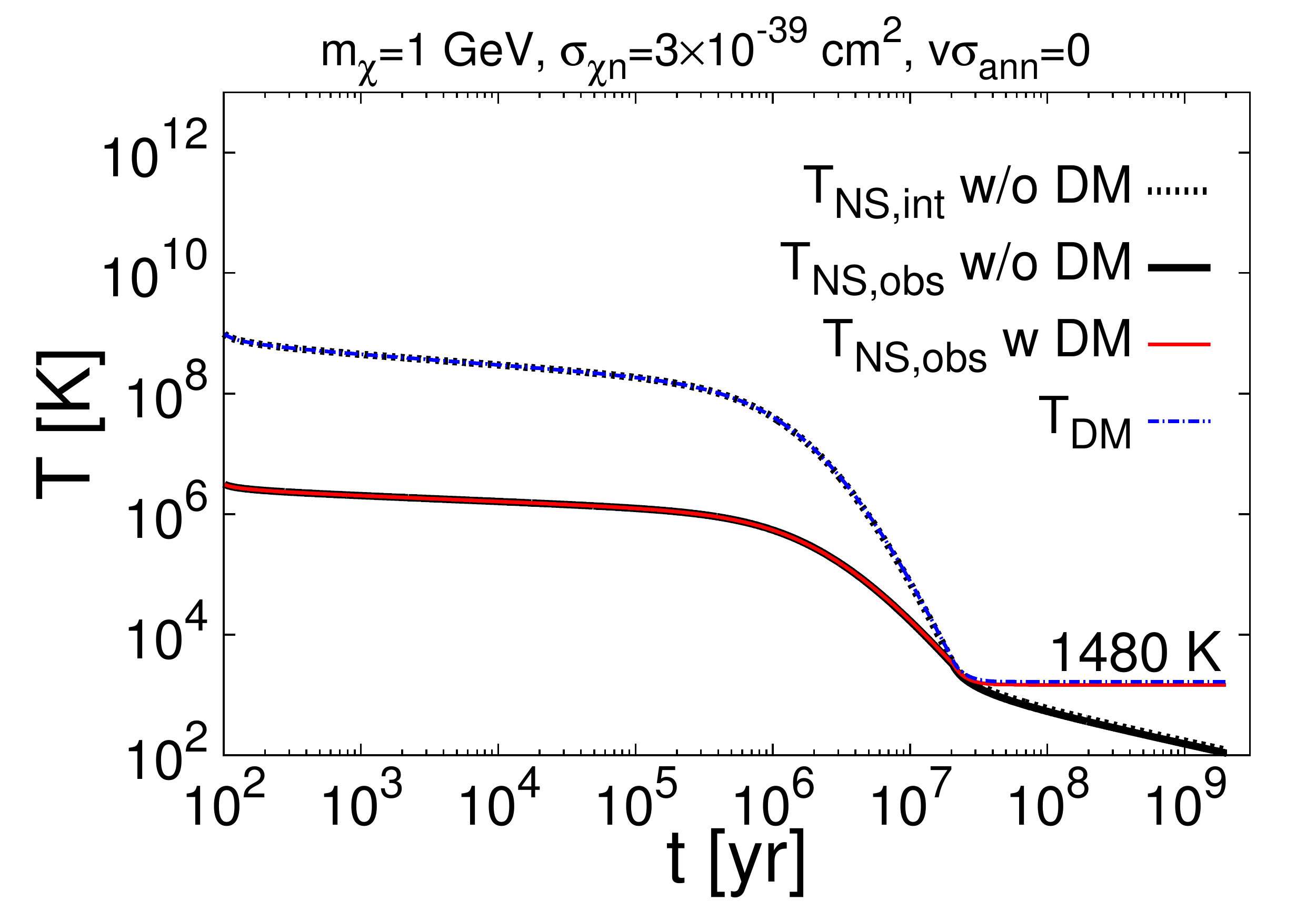}
\includegraphics[height=2.9in,angle=270]{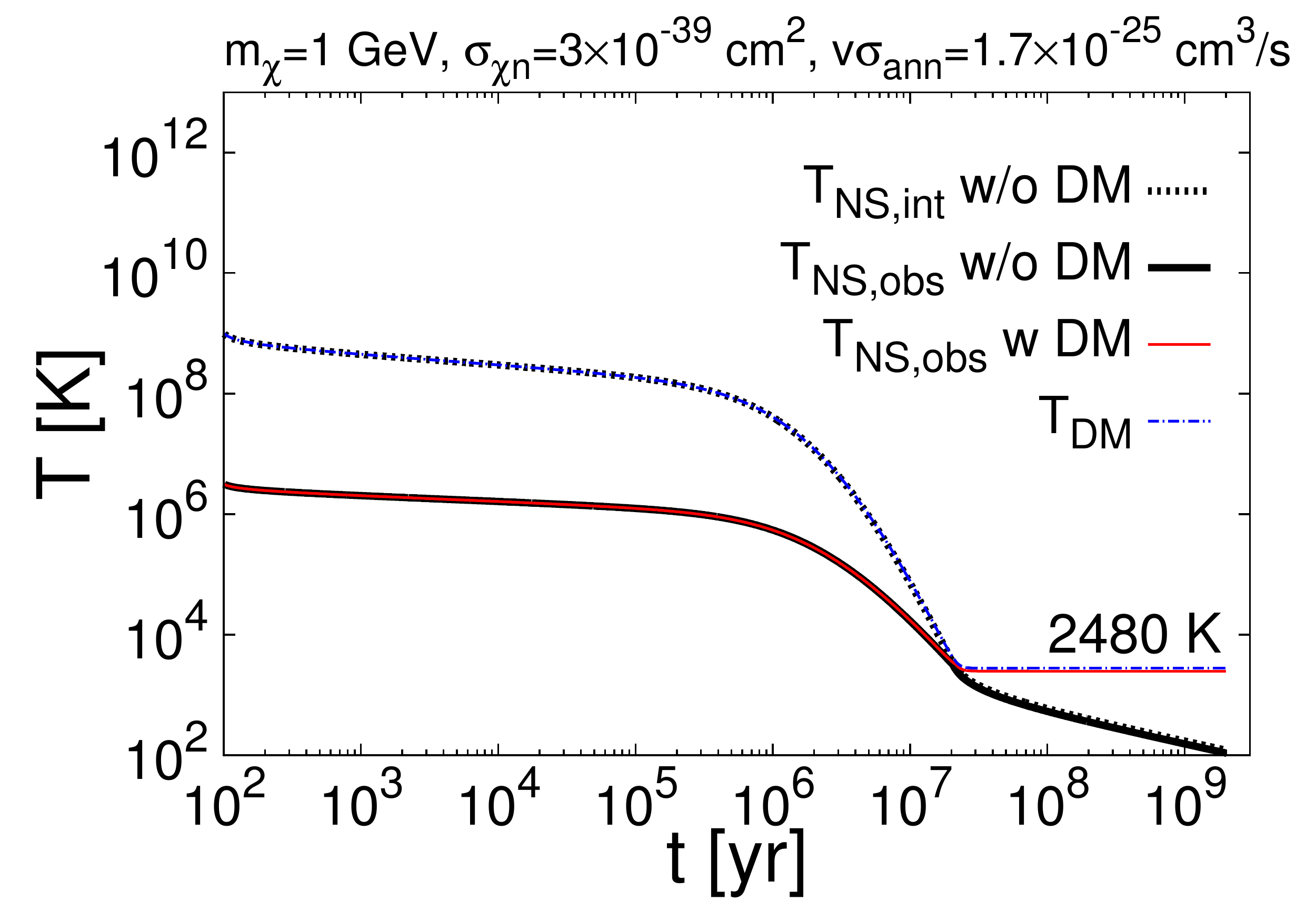}
\caption{\small \label{fig:T_DM}
The time evolution of NS temperatures including DM heating. The left panel does not have a contribution from DM annihilation, and the right panel does.
}
\end{figure}

In the case of kinetic heating, if the capture rate is at the geometric limit, the observed (surface) temperature increases to 1480~(1660)~K after
the photon emission and DM kinetic heating processes attain equilibrium,
$
L_\gamma|_{T_{\rm sur}=1660\,{\rm K}} = C_c|_{\rm geom}\langle E_R \rangle\,.
$
The left panel of Fig.~\ref{fig:T_DM} shows that  $T_{\rm obs}$ flattens out at 1480~K after
$5\times 10^7$~yrs.
 
 DM annihilation consumes the entire DM mass to heat up the NS, and if the annihilation rate is high enough, 
photon emission and DM heating reach equilibrium earlier. The observed (surface) temperature increases to 2480~(2780)~K,
when the photon emission energy-loss rate
equals the sum of the DM kinetic and annihilation heating rate:
$
L_\gamma|_{T_{\rm sur}=2780\,{\rm K}} = C_c|_{\rm geom}(\langle E_R \rangle+m_\chi)\,; 
$
see the right panel of Fig.~\ref{fig:T_DM}.
The surface temperature $T_{\rm sur}$ saturates at 2780~K, when the DM annihilation rate equals the
DM capture rate, i.e.,
$
N^2_{\rm DM} C_a|_{\rm sat} \simeq C_c\,.
$
Estimating $N_{\rm DM}$ by multiplying $C_c=C_c|_{\rm geom}$ with the typical age of an old NS, $5\times 10^8$~yr, 
we find the saturating DM annihilation cross section to be
$v_{\rm DM}\sigma^{\rm ann}_{\bar\chi \chi}|_{\rm sat}\simeq 10^{-39}\,{\rm cm^3/s}$.
Clearly, increasing $v_{\rm DM}\sigma^{\rm ann}_{\bar\chi \chi}$ above
 $v_{\rm DM}\sigma^{\rm ann}_{\bar\chi \chi}|_{\rm sat}$ does not increase $T_{\rm obs}$.

In general, the value of  $v_{\rm DM}\sigma^{\rm ann}_{\bar\chi \chi}|_{\rm sat}$ depends
 on $C_c$ and $\sigma^{\rm elastic}_{{\rm DM}-n}$.
For example, consider a smaller capture rate, $C_c= 10^{-4}\times C_c|_{\rm geom}$.
Without the heating from DM annihilation, 
the equilibrium condition, $L_\gamma|_{T_{\rm sur}=170\,{\rm K}}=C_c\langle E_R \rangle$,
gives a final NS surface temperature $T_{\rm sur}=170\,{\rm K}$.
Including DM annihilation increases the surface temperature to $T_{\rm sur}=280\,{\rm K}$ using the criterion,
$L_\gamma|_{T_{\rm sur}=280\,{\rm K}}=C_c(\langle E_R \rangle+m_\chi)$.
In this case,
$v_{\rm DM}\sigma^{\rm ann}_{\bar\chi \chi}|_{\rm sat}\simeq 10^{-35}\,{\rm cm^3/s}$.

In the neutron dark decay model, the trapped DM $\bar{\chi}$ 
can annihilate with 
the neutron or $\chi$ from neutron conversion to provide additional heating.
%from Eq.~(\ref{eq:ann_heating}), 
The observed (surface) temperature can reach 3100 (3440)~K,
if the photon emission energy-loss rate
equals the sum of the DM kinetic and annihilation heating rates:
$
L_\gamma|_{T_{\rm sur}=3440\,{\rm K}} = C_c|_{\rm geom}(\langle E_R \rangle+2m_\chi)\,.
$

\bigskip

\section{Neutron dark decay model}
\label{sec:neutron_dark}

The defining feature of the neutron dark decay model is that the neutron decays to dark sector particles $\chi$
and $\phi$. In the low energy limit, this can be described as a mixing between the neutron and the Dirac particle $\chi$, 
which could serve as DM.
However, since the DM particle is a Dirac fermion,
either $\chi$ or $\bar{\chi}$ could be DM, with different interactions with the neutron. Only $\bar{\chi}$ can annihilate with the neutron, 
and only $\chi$ is produced from neutron conversion. 
We separately discuss the phenomenologies of NS heating 
for these two cases.

\subsection{Model and NS equation of state}

The interaction terms in the model 
are~\cite{Fornal:2018eol,Grinstein:2018ptl}
\begin{eqnarray}
\mathcal{L}\supset && \lambda_q \epsilon^{ijk}\bar{u}^c_{Li}d_{Rj} \Phi_k
                  +\lambda_\chi \Phi^{*i}\bar{\tilde{\chi}}d_{Ri}
                  +\lambda_\phi \bar{\tilde{\chi}}\chi \phi 
                  + \mu H^\dagger H \phi + g_\chi \bar{\chi} \chi \phi
                  + {\rm h.c.}\,,
                  \label{Lag}
\end{eqnarray}
where the heavy scalar $\Phi=(3,1)_{-1/3}$ 
(color triplet, weak singlet, hypercharge -1/3) has mass above a TeV,
and two Dirac fermions $\tilde{\chi}$ and $\chi$, and a scalar $\phi$, are SM singlets
The baryon number assignments for $\Phi,\tilde{\chi},\chi,\phi$ 
are $-2/3,1,1,0$, respectively.{\footnote{The asymmetry between $\chi$ and $\bar{\chi}$ may originate as in models of
asymmetric dark matter~\cite{Nussinov:1985xr,Kaplan:2009ag}.
Since $\chi$ has the same baryon number as the neutron, chemical equilibrium in the early universe may relate the DM asymmetry to the baryon asymmetry. 
In asymmetric dark matter models, the DM particle has a GeV mass 
to reproduce the observed relic abundance.}
The annihilation process $\bar{\chi}\chi \to \phi \phi$ 
produces the observed DM relic abundance if the coupling $\lambda_\phi\simeq 0.04$.
The first three interaction terms
allow the decay $n \to \chi \phi$, 
which makes the NS unstable~\cite{McKeen:2018xwc}.
Including the Higgs portal and the  $g_\chi \bar{\chi} \chi \phi$ coupling, 
induces a repulsive 
$\chi$-neutron interaction, which causes
the energy density to increase when converting a neutron into $\chi$, 
so that the neutron becomes stable inside a NS~\cite{Grinstein:2018ptl}.
Then the interaction $g_n \bar{n}n\phi$ 
is generated from the Higgs portal interaction through the pion 
with
\begin{eqnarray}
g_n=\frac{\mu \sigma_{\pi n}}{m^2_h}\,,
\end{eqnarray}
where $\sigma_{\pi n}=370$ MeV and Higgs mass $m_h=125$ GeV.
 
Constraints from rapid red giant star cooling~\cite{Heeck:2014zfa} require $|g_n|\lesssim 10^{-14}$.
The sufficient condition to stabilize the NS is~\cite{Grinstein:2018ptl}
\begin{eqnarray}
z\equiv \frac{m_\phi}{\sqrt{|g_\chi g_n|}}\lesssim 71~\text{MeV} \,,
\end{eqnarray}
which puts the NS in the neutron phase, and no $\chi$ is produced.
Then the NS mass can reach two solar masses with central density of $6 n_0$.
For very light $\phi$, the choice,
$m_\phi \simeq 0.1~{\rm eV}$,
$g_\chi\simeq 4\times 10^{-4}$, and $\mu \simeq -0.4~{\rm eV}$, gives $z\simeq 50$ MeV to stabilize the NS, 
and also provides DM self-scattering cross sections of 
$0.1~{\rm cm^2/g} \lesssim \sigma/m_\chi\lesssim 1~{\rm cm^2/g}$,
which alleviates the tension between N-body simulations 
of collisionless cold DM and large scale structure observations~\cite{Grinstein:2018ptl}.  
However, if $m_\phi > 13$~eV, $g_n=-10^{-14}$, and $g_\chi \lesssim \sqrt{4 \pi}$,
$z$ can easily exceed 71~MeV.
Therefore, for heavier $\phi$, the NS is in a mixed phase,
and we must solve the equation of state (EoS) equation
to obtain the number densities, $n_\chi$ and $n_n$ in the NS.
In the mixed phase,
the NS can be stabilized by introducing a repulsive DM self-interaction,
and achieve a NS mass of about $2M_\odot$.

We solve the EoS equation as follows.
The energy density in a NS in a mixed phase is~\cite{Grinstein:2018ptl} 
\begin{equation}
\varepsilon(n_n,n_\chi)=\varepsilon_{\rm nuc}(n_n)+\varepsilon_\chi(n_\chi)
+\frac{n_\chi n_n}{2z^2}\,,
\end{equation}
where we assume $\chi$ is an ideal Fermi gas, and neutrons follow 
the EoS labeled
$V_{3\pi}+V_R$ in Ref.~\cite{Gandolfi:2011xu}, 
corresponding to moderately stiff EoSs
that incorporate 3-nucleon forces and have been fit to the results of a quantum Monte Carlo.
Then,
\begin{eqnarray}
&& \varepsilon_\chi=\frac{m^4_\chi}{8\pi^2}
\left[ x\sqrt{1+x^2}(1+2x^2)-\ln(x+\sqrt{1+x^2}) \right]
\pm \frac{n^2_\chi}{2z'^2} \,, \quad
x\equiv\frac{(3\pi^2 n_\chi)^{1/3}}{m_\chi}\,, \nonumber \\
&& \varepsilon_{\rm nuc}=ax'^\alpha + bx'^\beta\,,\quad 
x'\equiv \frac{n_n}{n_0}
\end{eqnarray}
with $a\,(b)=13.0\,(3.21)$ MeV, $\alpha\,(\beta)=0.49\,(2.47)$~\cite{Cline:2018ami}.
Here, $z'\equiv m_\phi/g_\chi$ comes from the DM self-interaction,
which if mediated by a scalar or vector boson results in an
attractive or repulsive force, respectively.
A repulsive DM self-interaction can be realized by introducing an
additional vector boson into the model; see Ref.~\cite{Cline:2018ami} for details
on the model construction.
Here, we simply fix the ratio of 
$z/z'=\sqrt{|g_\chi|/|g_n|}\simeq 2\times 10^5$, although in general, $z$ and $z'$ are two independent parameters. 
The equilibrium condition is
\begin{eqnarray}
0=\frac{\partial\varepsilon(n_F-n_\chi,n_\chi)}{\partial n_\chi}=
\mu_\chi(n_\chi)-\mu_{\rm nuc}(n_n)+\frac{n_F-2n_\chi}{2z^2}\,,
\end{eqnarray}
which is used to determine the $n$ and $\chi$ compositions of the NS.
The total Fermion number density satisfies $n_F=n_n+n_\chi$.
The neutron phase is determined by the condition
$\partial \varepsilon/\partial n_\chi |_{n_\chi=0}> 0$, which requires
that no $\chi$ be present, because introducing one $\chi$  increases the energy density.
On the other hand, the condition 
$\partial \varepsilon/\partial n_\chi |_{n_\chi=n_F}< 0$, transforms 
the entire NS into a $\chi$ star.
The mixed phase is defined by 
$\partial \varepsilon/\partial n_\chi |_{0< n_\chi < n_F} = 0$.
%for certain combinations between neutron and $\chi$.
The three phases are shown in the left panel 
of Fig.~\ref{fig:NS_EOS} in the $(z,n_F/n_0)$ plane.
The shading shows the density ratio 
$a_\chi\equiv n_\chi/(n_n+n_\chi)$, which is almost independent of $z$ for $z\gsim0.25$~GeV.
%The average number density $n_F\simeq 2.125\, n_0$ for NS considering in this work
%is indicated with the dashed horizontal line, and $\chi$ composes about 40\% of total number density of NS for $z>0.25$ GeV.
The minimal composition of $\chi$ 
occurs for $n_F\simeq  n_0$, in which case $\chi$ contributes about 30\% of the total number density.

\begin{figure}[t!]
\centering
\includegraphics[height=2.0in,angle=0]{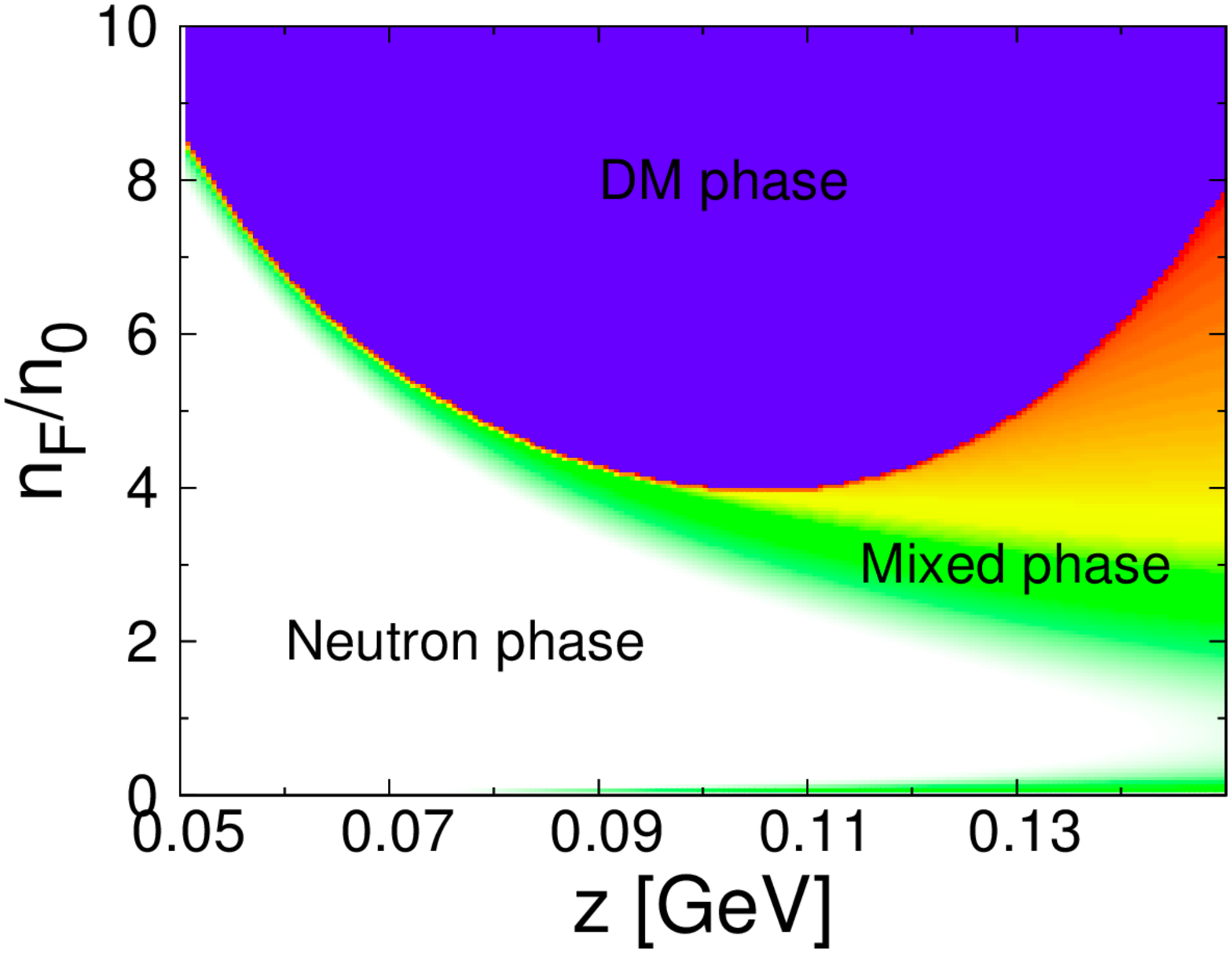}
\includegraphics[height=2.0in,angle=0]{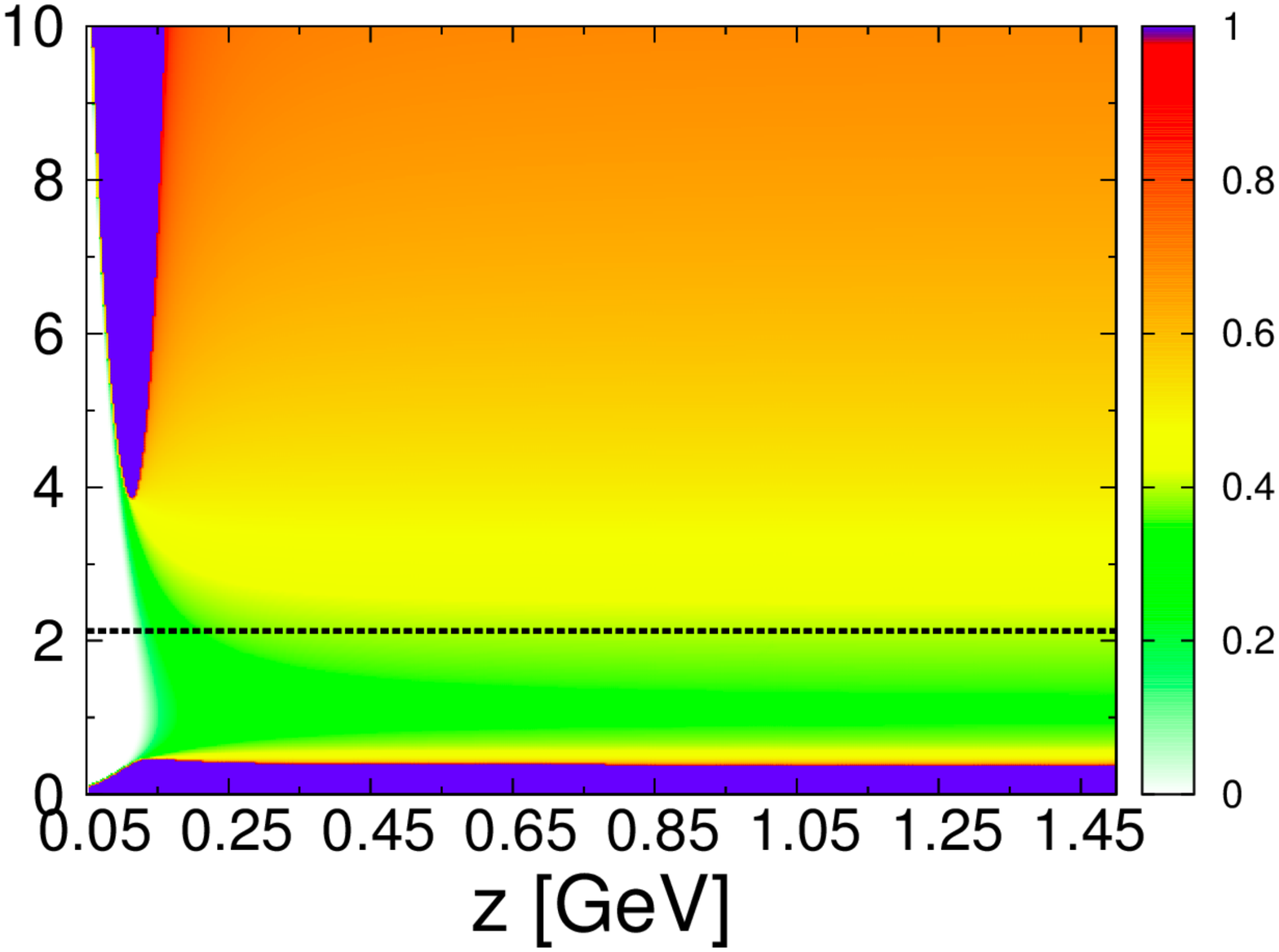}
\caption{\small \label{fig:NS_EOS}
The three phases of the NS. The right panel shows $n_F/n_0$ for a wider range in $z$. The shading
indicates $n_\chi/n_F$ for a given $n_F$.
For the NS we consider, $n_F\simeq 2.125\, n_0$, 
which is indicated by the dashed horizontal line.
In this case $\chi$ contributes about 40\% of the total number density of the NS.
}
\end{figure}

\begin{figure}[t!]
\centering
\includegraphics[height=2.0in,angle=0]{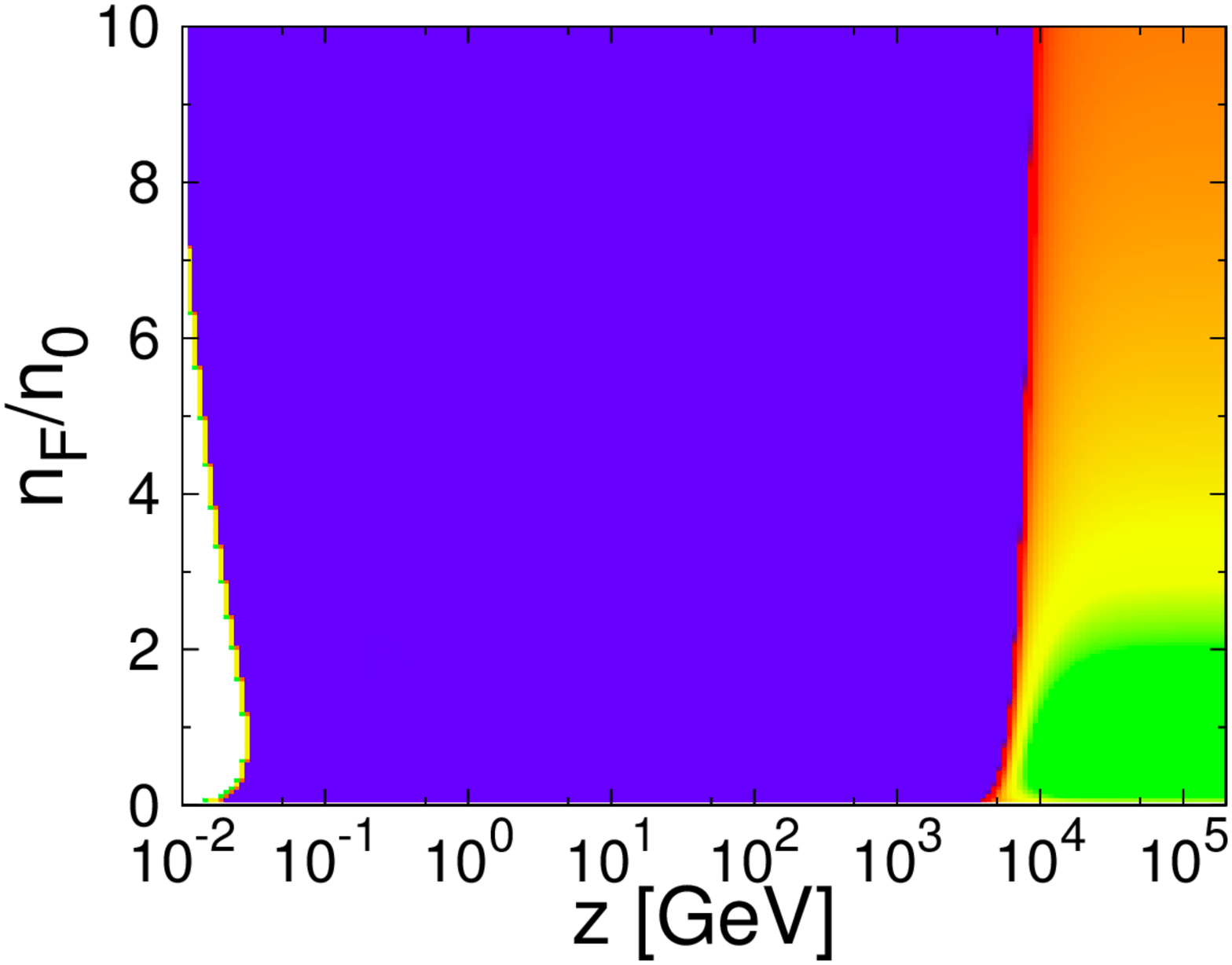}
\includegraphics[height=2.0in,angle=0]{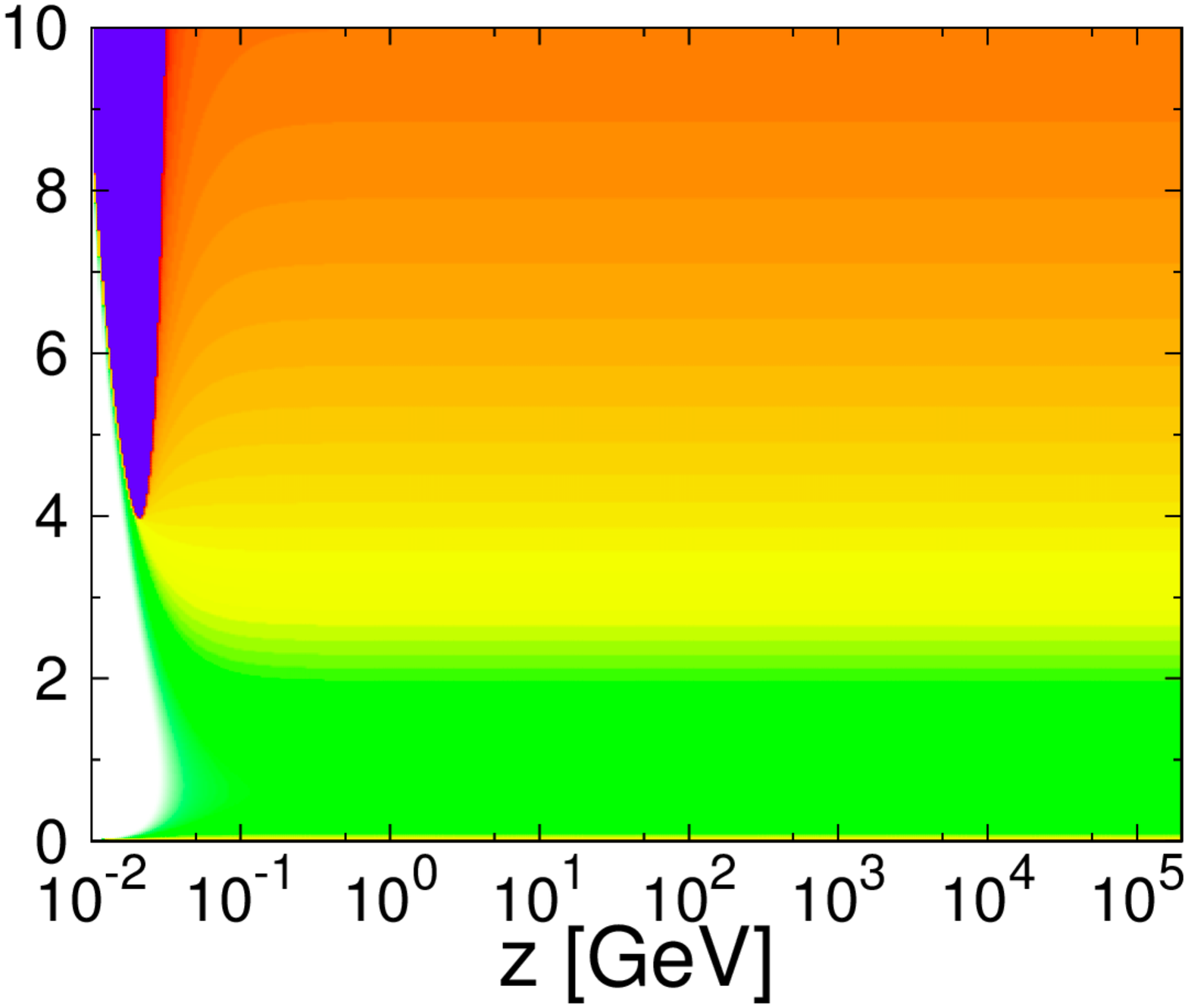}
\includegraphics[height=2.0in,angle=0]{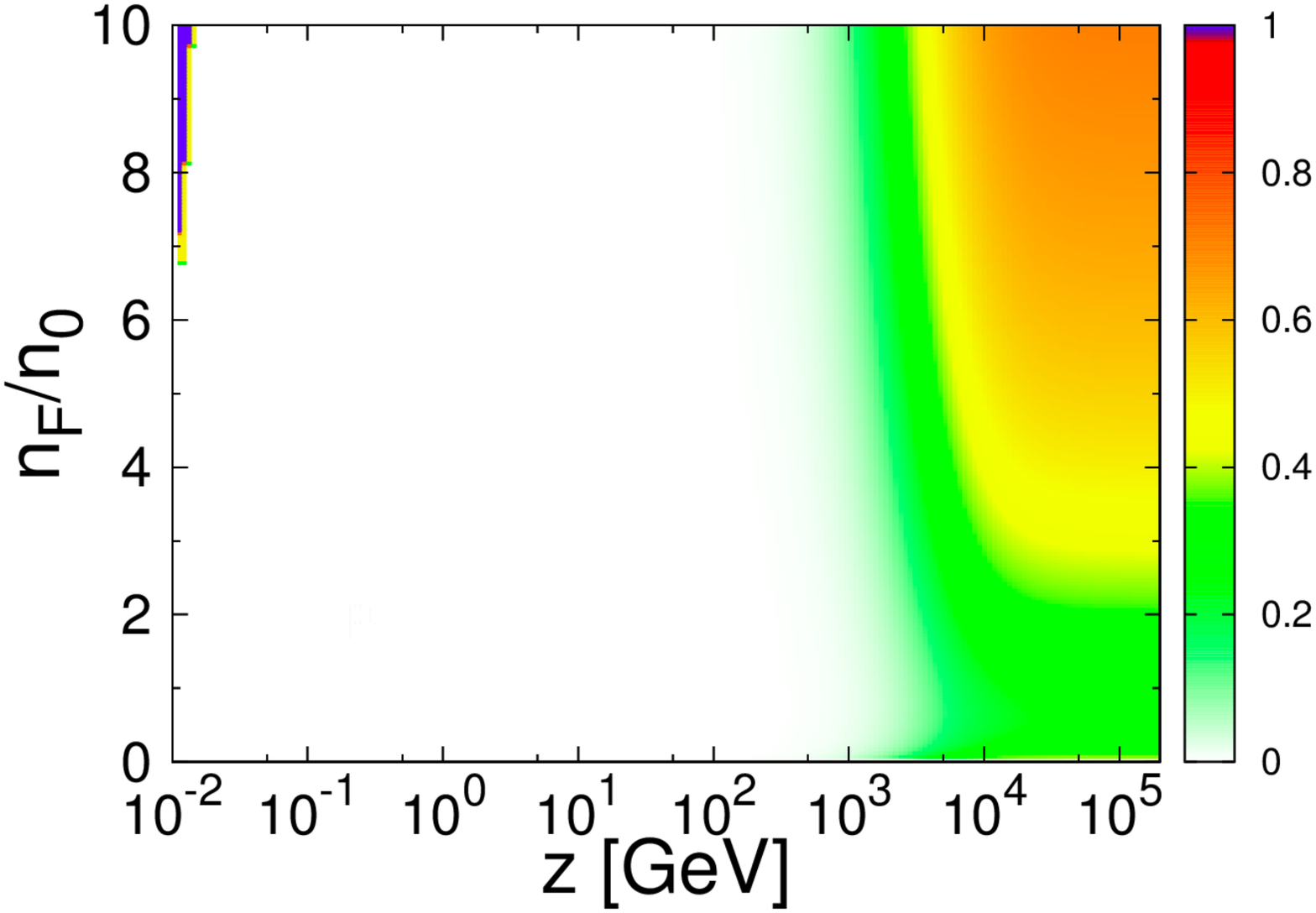}
\caption{\small \label{fig:NS_EOS_DM}
Same as Fig.~\ref{fig:NS_EOS}, but including DM self-energy and for $z/z'\simeq 2\times 10^5$.
Upper-Left panel: attractive DM self-energy $-\frac{n^2_\chi}{2z'^2}$.
Upper-Right panel: without DM self-energy.
Low-middle panel: repulsive DM self-energy $+\frac{n^2_\chi}{2z'^2}$.
}
\end{figure}

\begin{figure}[t!]
\centering
\includegraphics[height=1.5in,angle=0]{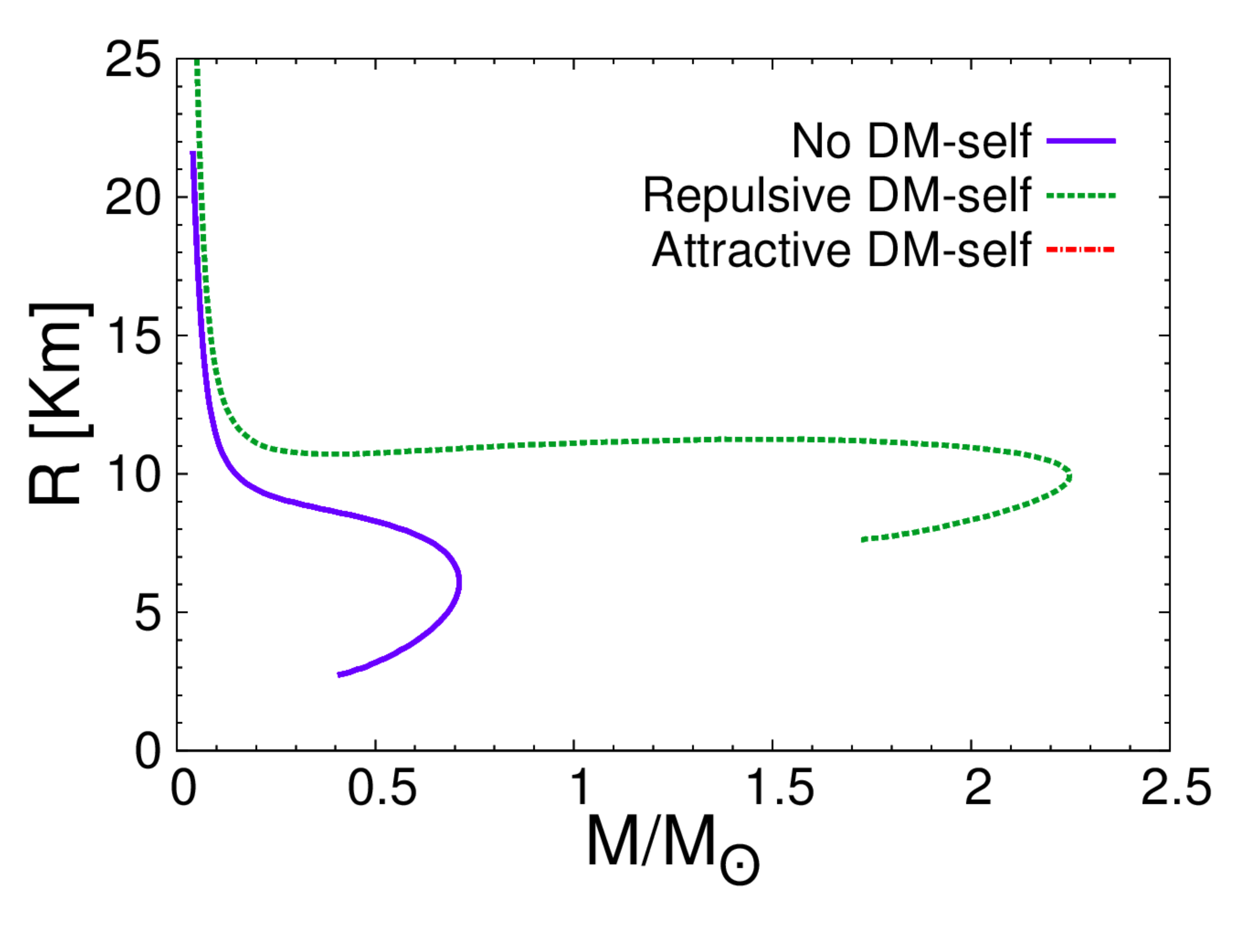}
\includegraphics[height=1.5in,angle=0]{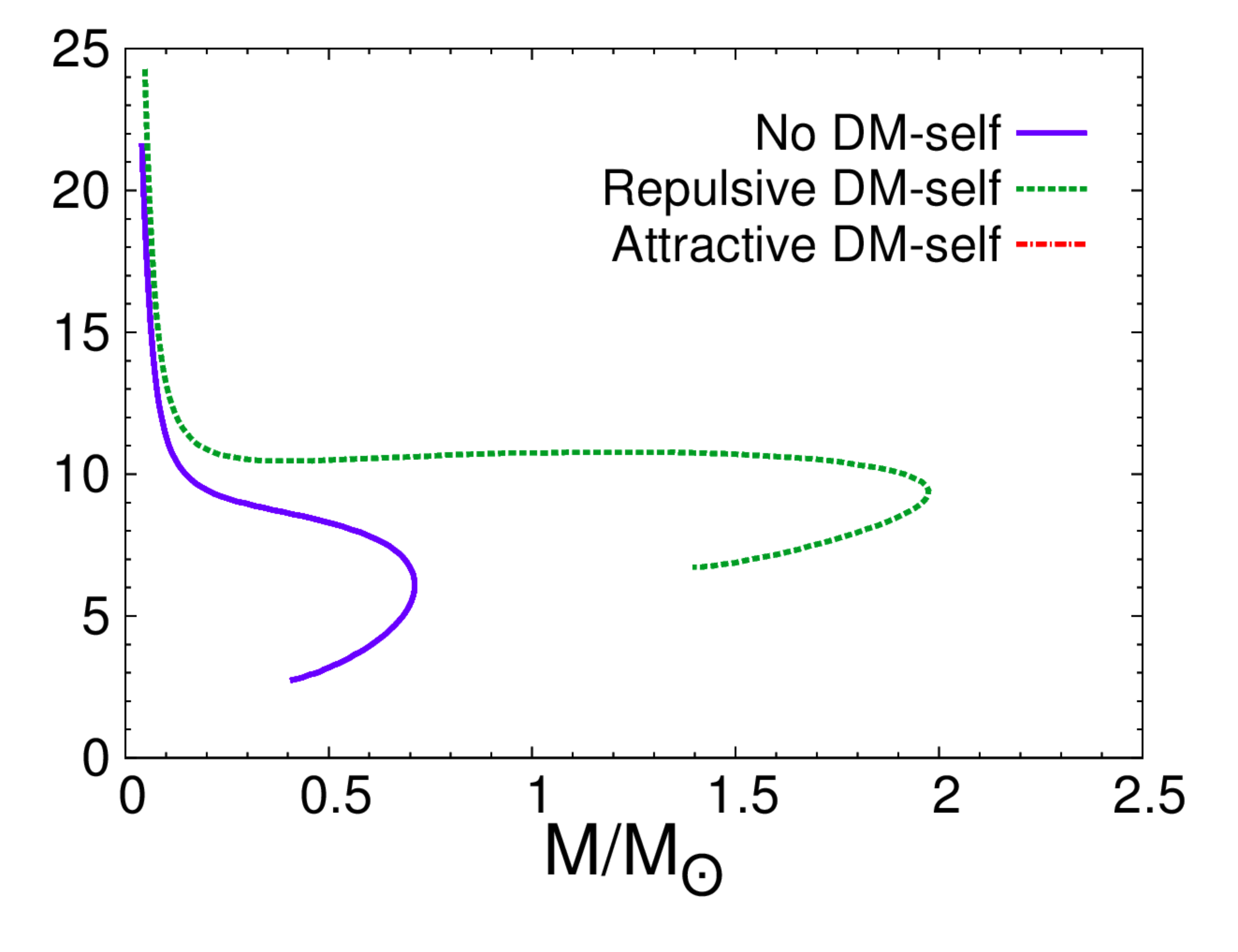}
\includegraphics[height=1.5in,angle=0]{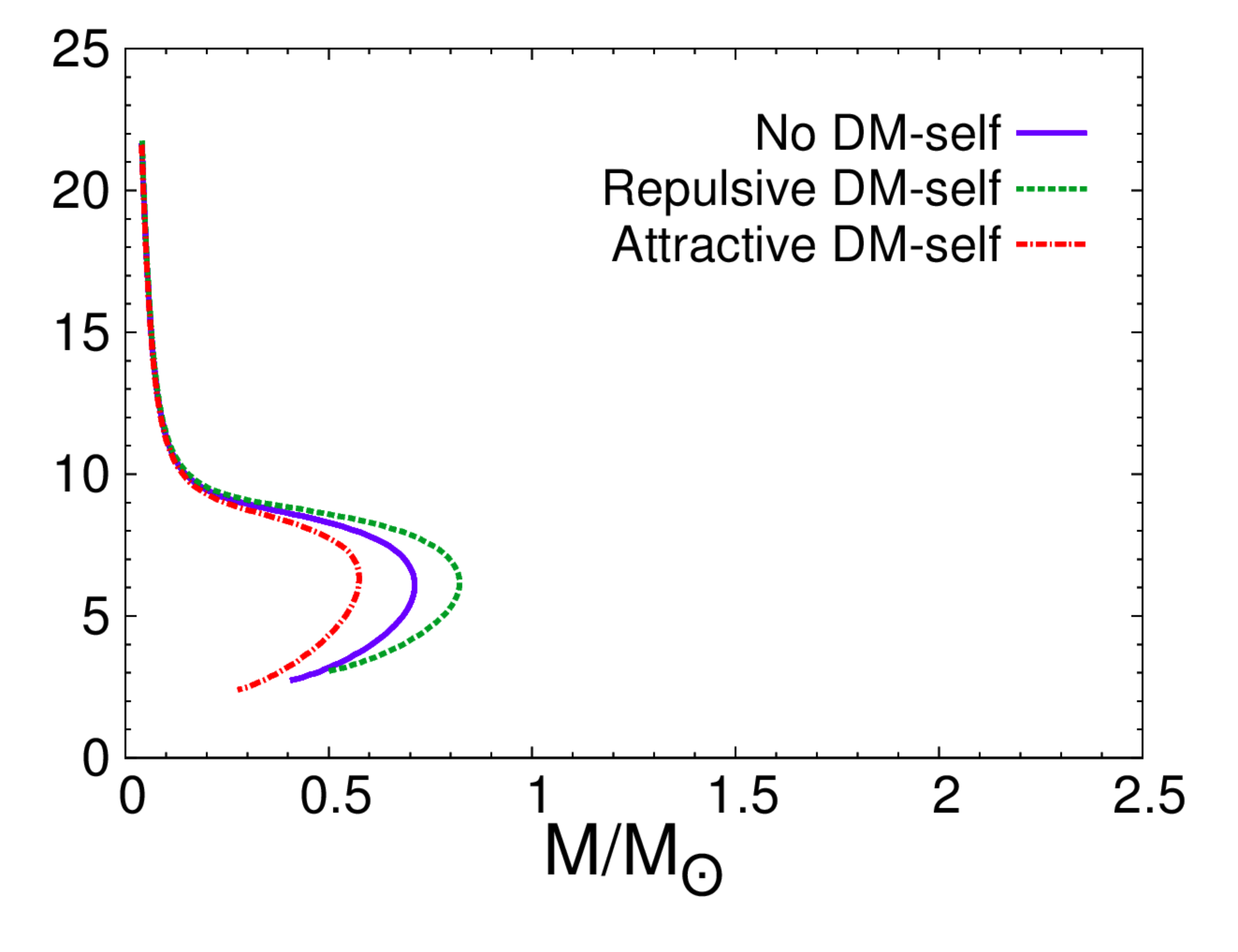}
\caption{\small \label{fig:TOV}
The NS mass for the neutron dark decay model for $z=10^{3},10^{4},10^{5}$~GeV in the left, middle, right panels, respectively,
with $z/z'\simeq 2\times 10^5$.
}
\end{figure}

The scenario with DM self-interactions is shown in Fig.~\ref{fig:NS_EOS_DM}.
The lower panel corresponds to repulsive DM self-interactions
which helps to stabilize the neutron star and 
 extends the neutron phase up to $z\simeq 10^3$~GeV. 
We also solve the 
Tolman-Oppenheimer-Volkoff equation~\cite{Douchin:2001sv} 
to check that neutron stars heavier than 2$M_\odot$ are obtainable.
From the correlation between total pressure $P=n^2_Fd(\varepsilon/n_F)/dn_F$ 
and $\varepsilon$, we find the relations between the NS mass 
and radius in Fig.~\ref{fig:TOV}.
From the left and middle panels we see that once $z'\lesssim 100$~MeV, 
the NS mass can be larger than 2$M_\odot$ for the repulsive case.
It is noteworthy that the NS in the repulsive case 
in the middle panel is in a mixed phase,
and can still reach 2$M_\odot$.

\subsection{DM-DM scattering cross section}

The DM self-scattering cross section arises from 
the $g_\chi \bar{\chi}\chi \phi$ and 
$\lambda_\phi \bar{\tilde{\chi}}\chi\phi$ terms in the Lagrangian.
The former is from the t-channel $\phi$ exchange diagram, 
while the later is generated from box diagrams with 
$\tilde{\chi}$ and $\phi$ in the loop.
Since $\lambda_\phi\simeq 0.04$ is 
much larger than $g_\chi\simeq 4\times 10^{-4}$, 
the loop-diagram contribution is comparable with the tree-level one.
Since a large fraction of the NS could be composed of $\chi$,
DM self-capture is crucial for NS heating.

The DM self-scattering 
cross section due to the $g_\chi \bar{\chi}\chi \phi$ term has been calculated in Ref.~\cite{Tulin:2012wi}.
%\begin{eqnarray}
%\sigma^{\rm eff}_{\chi \chi \to \chi \chi} \simeq \frac{4 \pi}{m^2_\phi}\beta^2 \ln %(1+\beta^{-1})\,,
%\end{eqnarray}
%when $\beta\equiv 2\alpha_\chi m_\phi/(m_\chi v^2) < 0.1$ 
%with $\alpha_\chi\equiv g^2_\chi/(4\pi)$.
The velocity-dependent cross section, 
which is inversely related to the fourth power of the velocity,
was proposed to solve the core-cusp problem.
During DM capture by a NS the typical DM velocity reaches 
$v\simeq 0.63 c$, which suppresses this cross section to
$\sigma^{\rm eff}_{\chi \chi \to \chi \chi}\simeq 8.0\times 10^{-40}~{\rm cm^2}$.
Thus, the DM self-scattering cross section from $g_\chi \bar{\chi}\chi$ becomes comparable to that from $\lambda_\phi \bar{\tilde{\chi}}\chi\phi$
 (via box diagrams), as we discuss below.

\begin{figure}[t!]
\centering
\includegraphics[height=2.2in,angle=0]{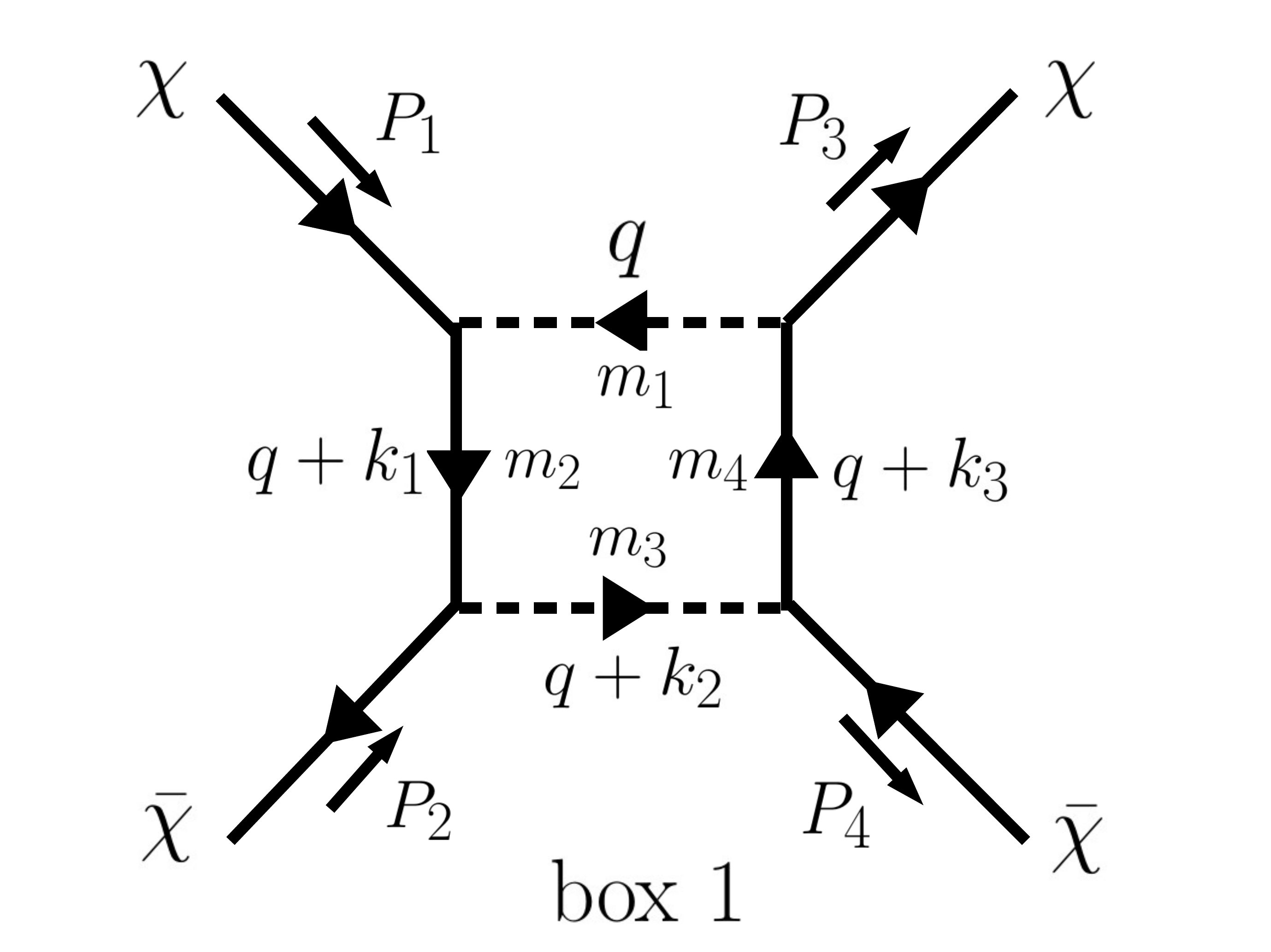}
\includegraphics[height=2.2in,angle=0]{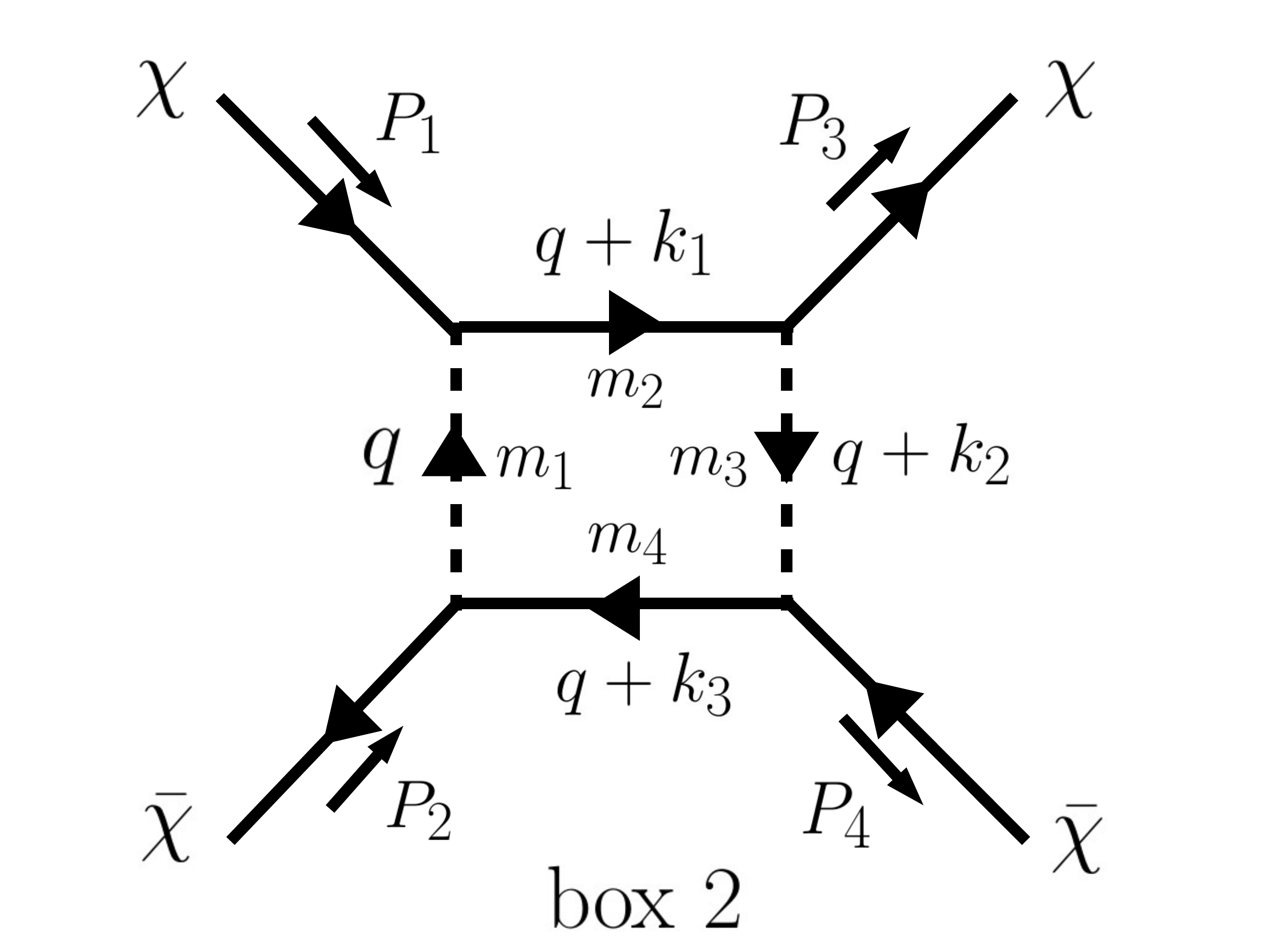}
\includegraphics[height=2.2in,angle=0]{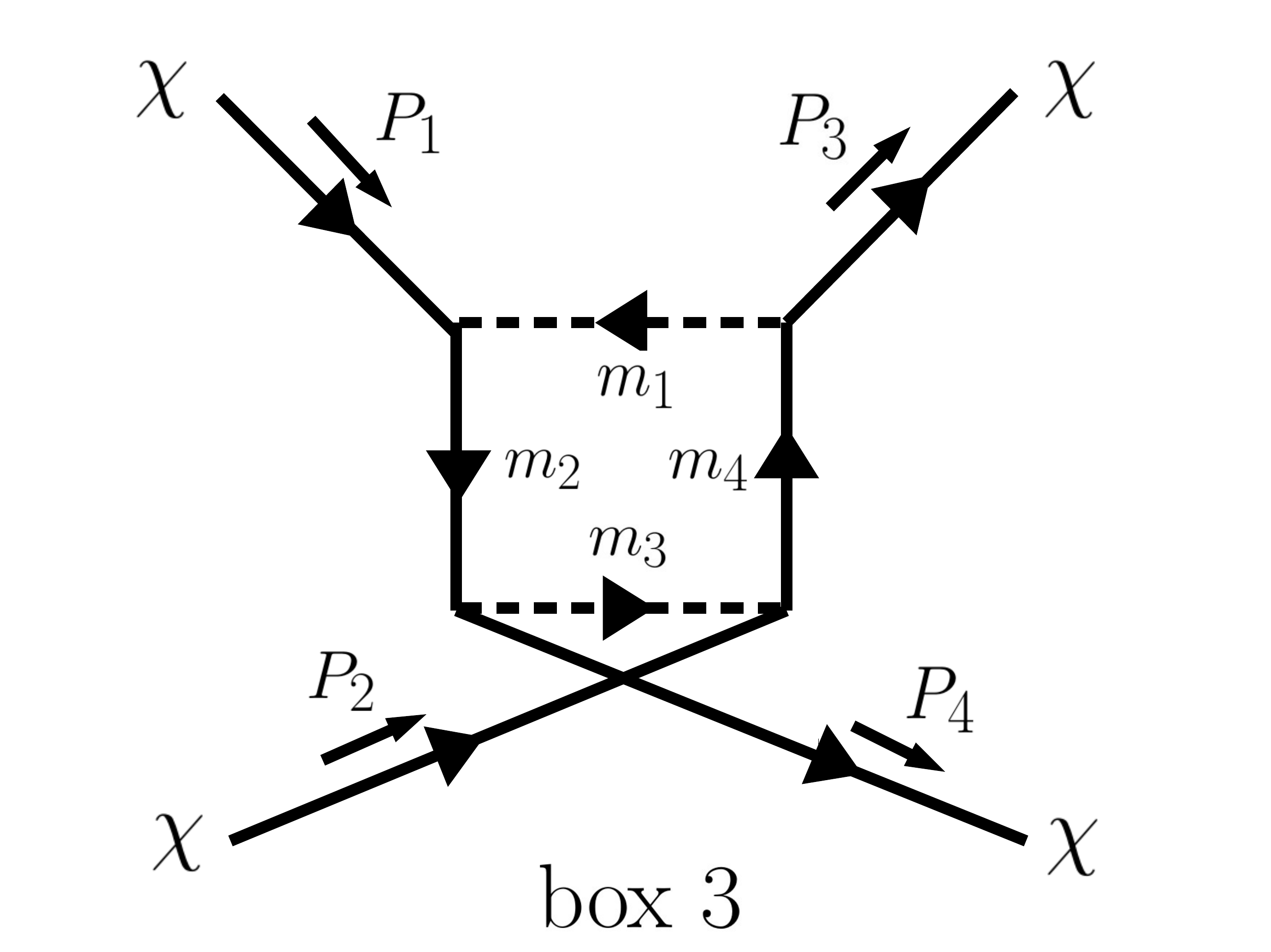}
\includegraphics[height=2.2in,angle=0]{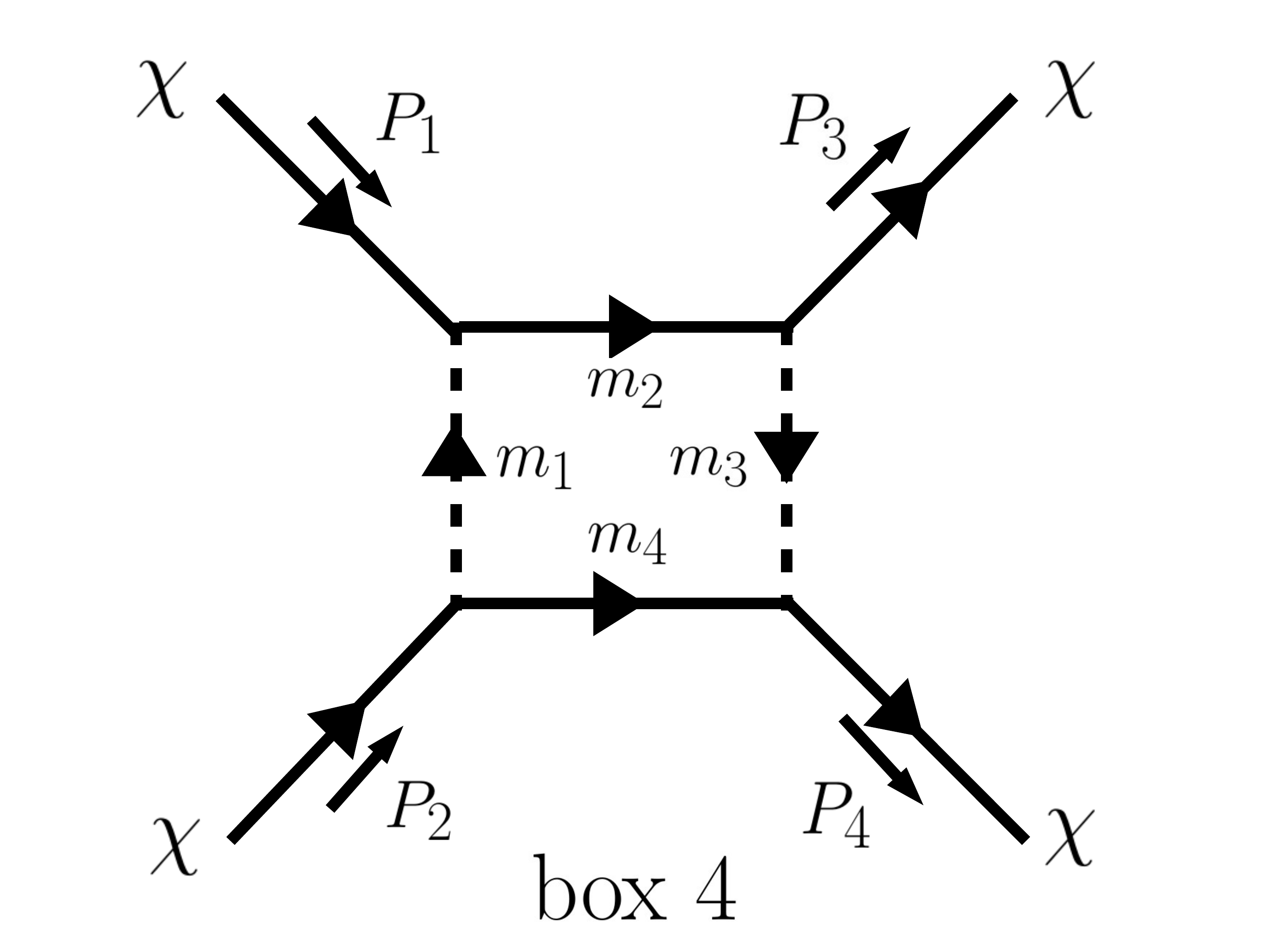}
\caption{\small \label{fig:diagram}
Box-1 and box-2 are for $\bar{\chi}\chi \to\bar{\chi}\chi$, while 
box-3 and box-4 are for $\chi\chi \to \chi\chi$, 
where $m_1=m_3=m_\phi$, $m_2=m_4=m_{\tilde{\chi}}$.
In box-1, $k_1=p_1$, $k_2=p_1+p_2$, $k_3=p_3$, 
while in box-2, $k_1=p_1$, $k_2=p_1-p_3$, $k_3=-p_2$.
}
\end{figure}

The DM self-scattering diagrams from the $\lambda_\phi \bar{\tilde{\chi}}\chi\phi$ term, 
 are shown in Fig.~\ref{fig:diagram}. The amplitudes for
$\chi(p_1) \bar{\chi}(p_2) \to \chi(p_3) \bar{\chi}(p_4)$
from box-1 and box-2 of Fig.~\ref{fig:diagram} are, respectively,
\begin{eqnarray}
\label{eq:box1}
i{\cal M}_{\rm box-1}= (\lambda^4_\phi) &&
\left\lbrace  [\bar{v}(p_2)\gamma_{\mu}u(p_1)][\bar{u}(p_3)\gamma_\nu v(p_4)]D^{\mu \nu}  \right.  \nonumber \\
+ && (m_\chi + m_{\tilde{\chi}}) [\bar{v}(p_2)\gamma_{\mu}u(p_1)]
[\bar{u}(p_3) v(p_4)]D^{\mu} \nonumber \\
+ && (m_\chi + m_{\tilde{\chi}}) [\bar{v}(p_2)u(p_1)]
[\bar{u}(p_3)\gamma_{\nu} v(p_4)]D^{\nu} \nonumber \\
+ && \left. (m_\chi + m_{\tilde{\chi}})^2 [\bar{v}(p_2)u(p_1)]
[\bar{u}(p_3) v(p_4)]D_0 
 \right\rbrace \,,
 \end{eqnarray}
\begin{eqnarray}
\label{eq:box-2}
i{\cal M}_{\rm box-2}= -(\lambda^4_\phi) &&
\left\lbrace  [\bar{v}(p_2)\gamma_{\mu}v(p_4)][\bar{u}(p_3)\gamma_\nu u(p_1)]D^{\mu \nu}  \right.  \nonumber \\
+ && (m_\chi + m_{\tilde{\chi}}) [\bar{v}(p_2)\gamma_{\mu}v(p_4)]
[\bar{u}(p_3) u(p_1)]D^{\mu} \nonumber \\
+ && (m_\chi + m_{\tilde{\chi}}) [\bar{v}(p_2)v(p_4)]
[\bar{u}(p_3)\gamma_{\nu} u(p_1)]D^{\nu} \nonumber \\
+ && \left. (m_\chi + m_{\tilde{\chi}})^2 [\bar{v}(p_2)v(p_4)]
[\bar{u}(p_3) u(p_1)]D_0 
 \right\rbrace \,,
\end{eqnarray}
where the relative minus sign arises from Fermi statistics. $D^{\mu \nu}$, $D^{\mu,\nu}$ and $D_0$ are loop integration functions defined in $LoopTools$\,\cite{Hahn:1998yk} as
\begin{eqnarray}
D_0 &=& \frac{\mu^{4-d}}{i \pi^{d/2}\gamma_\Gamma}
\int dq^d 
\frac{1}{[q^2-m^2_1][(q+k_1)^2-m^2_2][(q+k_2)^2-m^2_3][(q+k_3)^2-m^2_4]}\,, \nonumber \\
D^\mu &=& \frac{\mu^{4-d}}{i \pi^{d/2}\gamma_\Gamma}
\int dq^d
\frac{q^\mu}{[q^2-m^2_1][(q+k_1)^2-m^2_2][(q+k_2)^2-m^2_3][(q+k_3)^2-m^2_4]}\,, \nonumber \\
D^{\mu\nu} &=& \frac{\mu^{4-d}}{i \pi^{d/2}\gamma_\Gamma}
\int dq^d 
\frac{q^\mu q^\nu}{[q^2-m^2_1][(q+k_1)^2-m^2_2][(q+k_2)^2-m^2_3][(q+k_3)^2-m^2_4]}\,,
\end{eqnarray}
where 
$d=4-2\varepsilon$, $\gamma_\Gamma\equiv \frac{\Gamma^2(1-\varepsilon)\Gamma(1+\varepsilon)}{\Gamma(1-2\varepsilon)}$,
%from Gamma function $\Gamma(t)\equiv \int^{\infty}_0 x^{t-1}e^{-t} dt$, 
 and $\mu$ is the renormalization scale.
In order to match the Dirac spinors between box-1 and box-2, 
we use the Fierz transformation~\cite{Pal:2007dc}
\begin{eqnarray}
w_4 \bar{w}_3=\frac{1}{4}
\left[(\bar{w}_3 w_4)\mathbb{I}
+(\bar{w}_3 \gamma^\alpha w_4)\gamma_\alpha
+\frac{1}{2}(\bar{w}_3 \sigma^{\alpha \beta} w_4)\sigma_{\alpha \beta}
-(\bar{w}_3 \gamma^\alpha \gamma_5 w_4)\gamma_\alpha \gamma_5
+(\bar{w}_3 \gamma_5 w_4)\gamma_5
 \right]\,, \nonumber \\
\end{eqnarray}
where Dirac spinor $w$ represents either the $u$ or $v$ spinors. 
Then the crossing operation,  $p_2 \to -p_4$ and $p_4 \to -p_2$, 
yields the amplitude for the DM self-scattering cross section $\chi \chi \to \chi \chi$
from box-3 and box-4.

The box diagrams are significantly enhanced by the $D^{\mu \nu}$ loop function
when the scattering angle in the centre of mass frame 
approaches 
$\theta_{\rm cm}\simeq 0$ or $\pi$. This is due to the nearly massless mediator $\phi$.
Fortunately, neither collinear nor head on scattering 
contribute to the DM captured by DM inside the NS because the net trapped DM number remains the same
in both cases.
The energy transfer in a DM-DM collision is given by~\cite{Bell:2018pkk}
\begin{equation}
\frac{(1-\bar{B})m_\chi}{2\bar{B}+2\sqrt{\bar{B}}}(1-\cos\theta_{\rm cm})\,,
\end{equation}
where $\bar{B}\equiv 1-2GM/(c^2 R)$ for a NS of mass $M$ and radius $R$.
So, the collinear scattering ($\theta_{\rm cm}\simeq 0$) 
cannot slow down the incoming DM enough to be trapped by the NS.
On the other hand, head on scattering $\theta_{\rm cm}\simeq \pi$
exchanges the momenta of the two initial DM particles 
such that the incoming DM particle gets trapped and the target particle 
gets kicked out of the NS.

We define an effective DM self-scattering cross section,
which is relevant to the DM captured inside the NS:
\begin{eqnarray}
\sigma^{\rm eff}_{\chi \bar{\chi} \to \chi \bar{\chi}}\equiv 
\int^\pi_0 d\theta_{\rm cm} 
\frac{d\sigma_{\chi \bar{\chi} \to \chi \bar{\chi}}}{d\theta_{\rm cm}}
(1-\cos\theta_{\rm cm})(1+\cos\theta_{\rm cm})\,,
\end{eqnarray}
and similarly  for $\chi \chi \to \chi \chi$.
The
$(1-\cos \theta_{\rm cm})$ and $(1+\cos \theta_{\rm cm})$ factors are included
to suppress the phase space contributions from collinear and head-on scatterings, respectively~\cite{Raby:1987nb}.
These factors also cancel the infrared divergence 
in $d \sigma_{\chi \bar{\chi}\to \chi \bar{\chi}} / d \theta_{\rm cm}$ that
originates from the exchange of the light mediator $\phi$,
thereby rendering $\sigma^{\rm eff}_{\chi \bar{\chi} \to \chi \bar{\chi}}$ finite.
The cross sections in Fig.~\ref{fig:XXXX} are finite.
The loop-level contribution from $\lambda_\phi \bar{\tilde{\chi}}\chi\phi$ 
is comparable with the tree-level contribution from $g_\chi \bar{\chi}\chi \phi$
because $\lambda_\phi \gg g_\chi$.

\begin{figure}[t!]
\centering
\includegraphics[height=2.9in,angle=270]{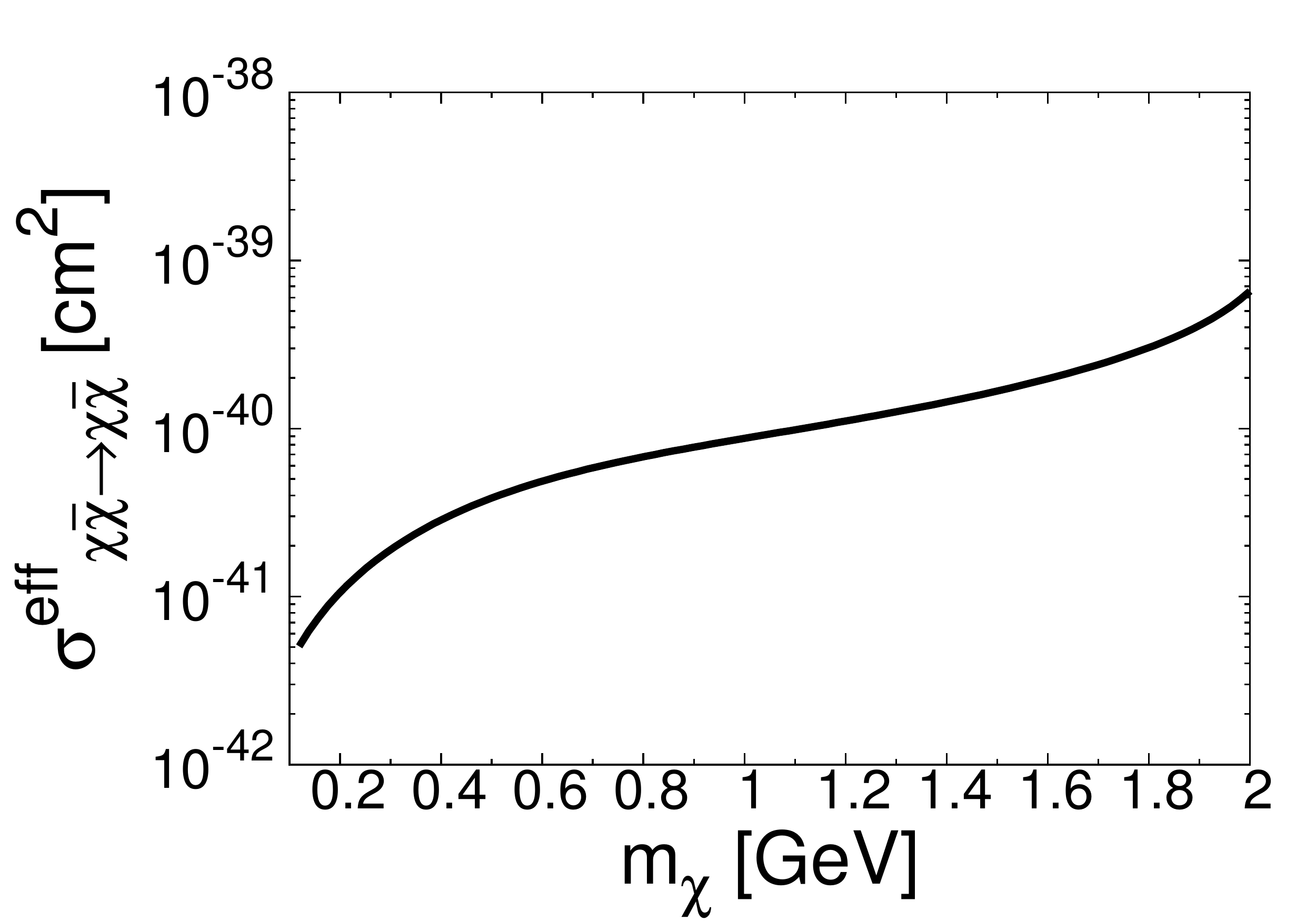}
\includegraphics[height=2.9in,angle=270]{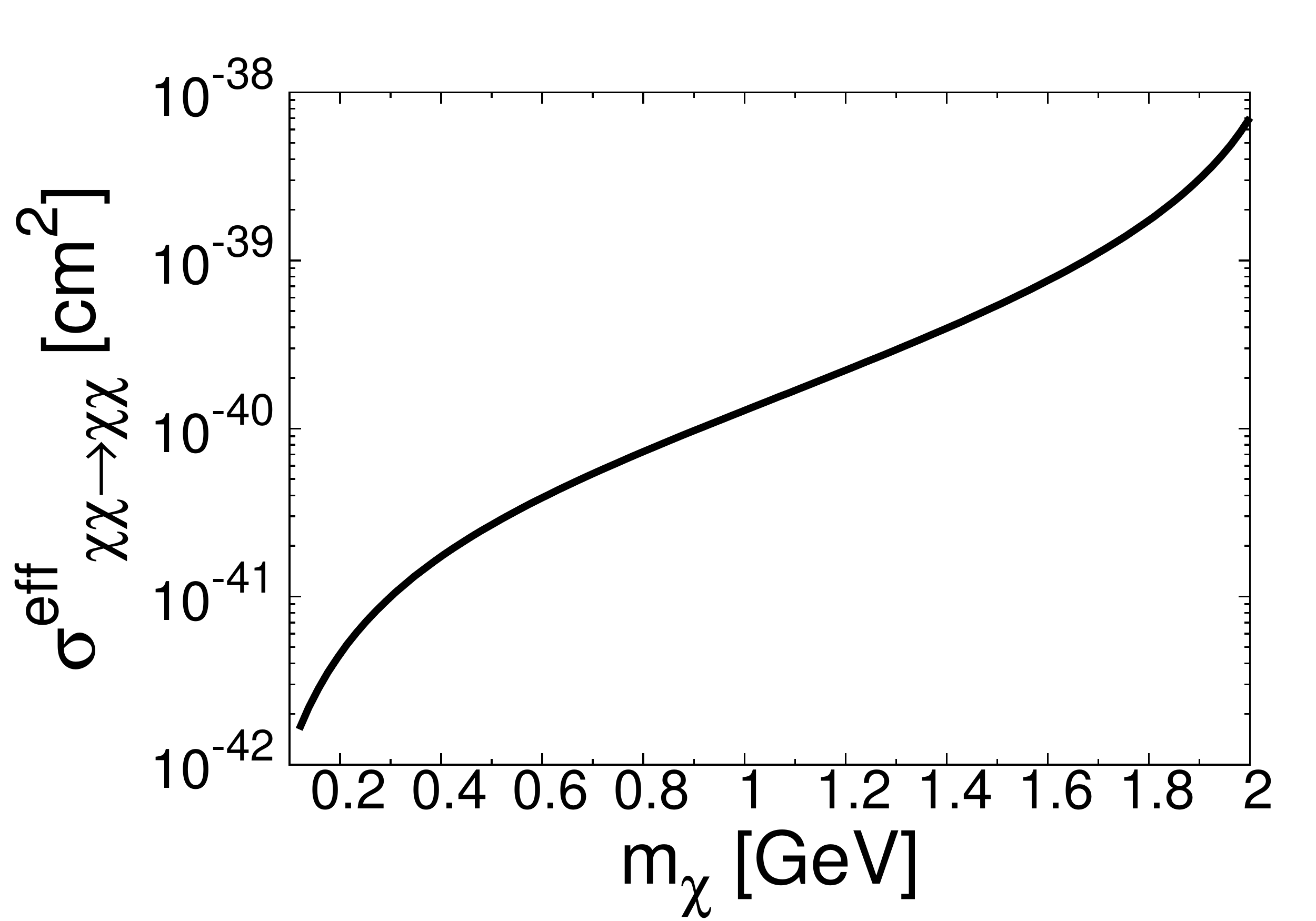}
\caption{\small \label{fig:XXXX}
The cross sections $\sigma^{\rm eff}_{\chi \bar{\chi} \to \chi \bar{\chi}}$
and $\sigma^{\rm eff}_{\chi \chi \to \chi \chi}$
from the two box diagrams of Fig.~\ref{fig:diagram}.
We set $\lambda_\phi=0.04$, $m_{\tilde{\chi}}=2$ GeV, $m_\phi=0.1$ MeV.
}
\end{figure}

\subsection{DM-neutron elastic scattering and annihilation cross sections}

At the GeV energy scale, the model can be described by an effective operator,
$\mathcal{L}\supset \varepsilon (\bar{n}\tilde{\chi}+\bar{\tilde{\chi}}n)$,
which mixes  $n$ and $\tilde{\chi}$ with mixing angle 
$\theta=\varepsilon/(m_n-m_{\tilde{\chi}})$.
$\theta \simeq \mathcal{O}(10^{-11}-10^{-12})$ accommodates the neutron lifetime anomaly.
Then the DM-neutron elastic scattering cross section is obtained 
from t-channel $\phi$ exchange, $\chi n \to \phi \to \chi n$:
\begin{eqnarray}
%&&\sigma^{\rm elastic}_{\bar{\chi} n}
%=\sigma^{\rm elastic}_{\chi n} {\color{blue}\lesssim 
%\theta^4 \sigma^{\rm elastic}_{\bar{n}p}\simeq \mathcal{O}(10^{-69})
%~{\rm cm^2}}\,, \nonumber \\
%
\sigma^{\rm elastic}_{\chi n \to \phi \to \chi n}=
\sigma^{\rm elastic}_{\bar{\chi} n \to \phi \to \bar{\chi} n}
\simeq \mathcal{O}(10^{-60})
~{\rm cm^2}\,. \nonumber 
\end{eqnarray}
%{\color{red}
%Or from the simple estimation using the low energy 
%$\chi\text{-}n$ scattering length scale, it gives 
%$\sigma^{\rm elastic}_{\chi n}\simeq 2\times 10^{-54}~{\rm %cm^2}$\,\cite{Baym:2018ljz}.
%}
%

%where we assume $\sigma_{\bar{n}p}=\sigma_{\bar{n}n}$.
%And it is similarly for the $\bar{\chi}-n$ annihilation cross section 
%$\sigma^{\rm ann}_{\bar{\chi}n}$.
%
%Since the $\bar{n}p$ or $\bar{n}n$ scatting are mainly through the pion exchanging,
%the calculations are non-perturbative.
%For Alternative, we use the experimental approaches to obtain these cross sections.
%According to the experimental values~\cite{Iazzi:2000rk,Bertin:1997gn}, 
%the antineutron-proton elastic scattering cross section is 
%$\sigma^{\rm elastic}_{\bar{n}p}\simeq 100~[\rm mb]$
%at the antineutron momentum around 400 MeV. 
%which corresponds to the typical kinetic energy of 1 GeV DM falling into a NS.
For $\sigma^{\rm ann}_{\bar{\chi} n}$, the dominant annihilation mode is 
$\phi+{\rm multipion}$, which depends on $m_\chi,m_{\tilde{\chi}},m_\phi$.
The detailed calculations in Ref.~\cite{Keung:2019wpw} give
\begin{eqnarray}
 \sigma^{\rm ann}_{\bar{\chi} n} (v/c)\simeq \mathcal{O}(10^{-50}-10^{-54})
~{\rm cm^2}\,.\nonumber
\end{eqnarray}
Both $\sigma^{\rm elastic}_{\chi n}$ and $\sigma^{\rm ann}_{\bar{\chi} n}$ contribute negligibly to NS heating since $\sigma^{\rm elastic}_{\bar{\chi}n}\ll \sigma_{\rm crit}$ 
and $\sigma^{\rm ann}_{\bar{\chi} n}(v/c)\ll \sigma^{\rm ann}_{\bar{\chi} \chi}(v/c)$.
Therefore, in the following calculations, 
we conservatively fix $\sigma^{\rm elastic}_{\bar{\chi}n}=0$ and 
$\sigma^{\rm ann}_{\bar{\chi} n}(v/c)=10^{-54}~{\rm cm^2}$ 
to estimate NS heating.

\bigskip

\subsection{Results}
\label{sec:modelII_results}

The salient feature of this model is that
neutrons can convert into $\chi$ inside the NS, 
which makes the NS composed of $n$ and $\chi$ in most of the
interesting parameter space.
Then the DM self-scattering cross sections from the box diagrams in Fig.~\ref{fig:diagram}, 
that are significantly larger than the critical cross section $\sigma_{\rm crit}$,
enhance the capture rate above the geometric limit.
Consequently, the NS can be heated up to 1500~K.
If further $\bar{\chi}-n$ and $\bar{\chi}-\chi$ annihilations are allowed,
the NS temperature might reach 3100~K depending on the 
final state particles from annihilation. 

We are interested in the parameter regions which can explain 
the neutron lifetime anomaly. The masses $m_\chi$, $m_\phi$, 
and $m_{\tilde{\chi}}$ in this model need to satisfy the relations~\cite{Fornal:2018eol}
\begin{eqnarray}
\label{eq:parameter}
&& 937.992~{\rm MeV} < m_\chi+m_\phi < 939.565~{\rm MeV}\,, \nonumber \\
&& 937.992~{\rm MeV} < m_{\tilde{\chi}}\,, \nonumber \\
&& |m_\chi-m_\phi|< m_p+m_e = 938.783081~{\rm MeV}\,. 
\end{eqnarray}
 We choose three benchmark points of Ref.~\cite{Keung:2019wpw}, 
\begin{eqnarray}
{\bf P1}: ~~&& (m_\chi,m_\phi,m_{\tilde{\chi}})
=(937.992,0,937.992)  \nonumber \\
{\bf P2}: ~~ && (m_\chi,m_\phi,m_{\tilde{\chi}})
=(937.992,0,2m_n)  \nonumber \\
{\bf P3}: ~~ && (m_\chi,m_\phi,m_{\tilde{\chi}})
=(939.174,0.391,940.000)\,,  \nonumber 
\end{eqnarray}
within the region.
We fix $\lambda_\phi=0.04$ to give the correct 
DM relic density~\cite{Fornal:2018eol}, and $g_\chi=4\times 10^{-4}$ 
to alleviate the core-cusp problem~\cite{Grinstein:2018ptl}.

Note that the light mediator $\phi$ is not stable and decays to diphotons by mixing with the SM Higgs via
the $\mu H^\dagger H\phi$ term in Eq.~(\ref{Lag}). Also, because of its tiny mixing with the SM Higgs, $\phi$ decouples from the
primordial plasma before neutrino decoupling. Thus, $\phi$ does not contribute
to the effective number of relativistic degrees of freedom in the early universe. 

For the neutron dark decay model, the DM can be either $\bar{\chi}$ or $\chi$,
so we separately discuss these cases below.

\subsubsection{$\chi$ is DM}

\begin{figure}[t!]
\centering
\includegraphics[height=1.5in,angle=0]{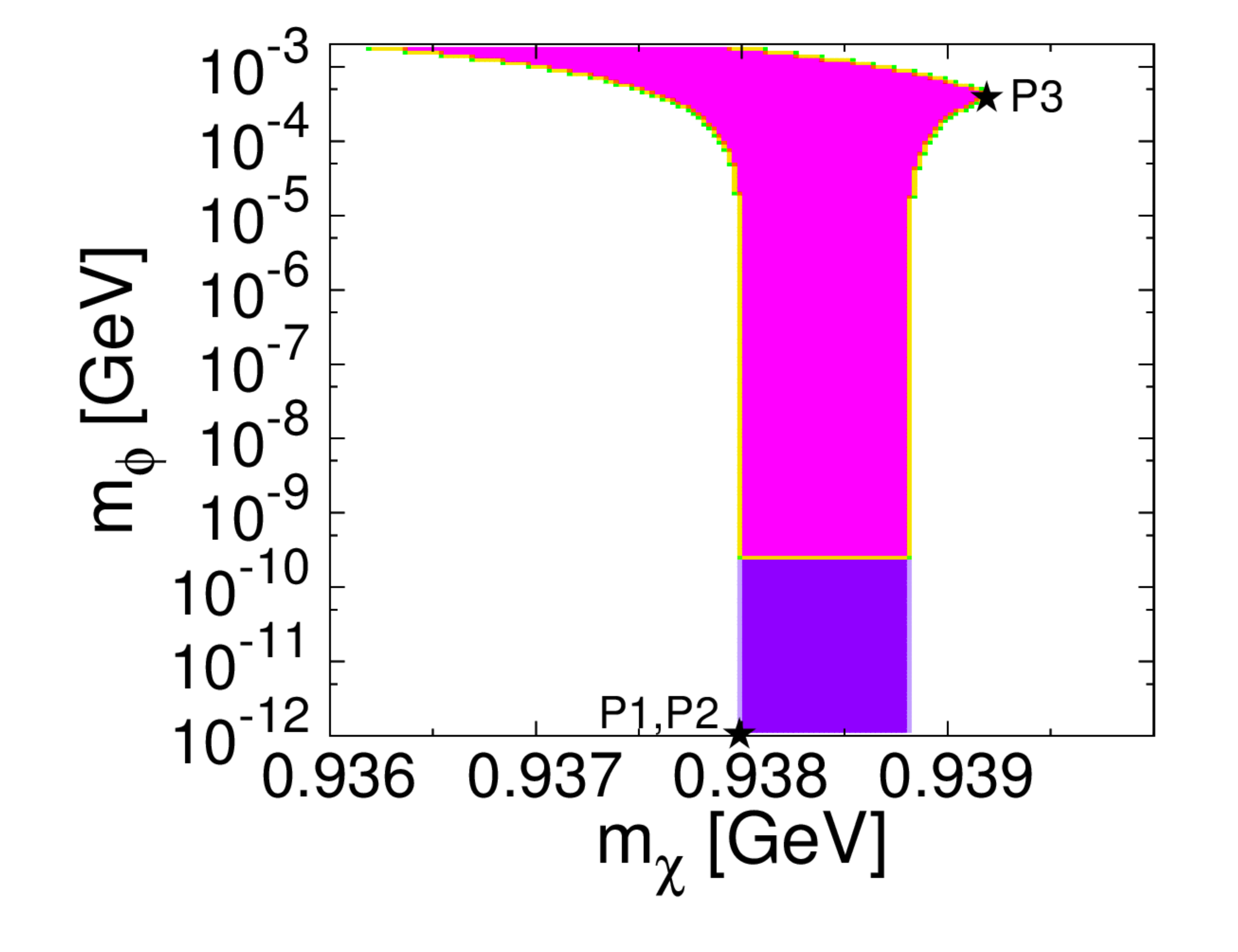}
\includegraphics[height=1.5in,angle=0]{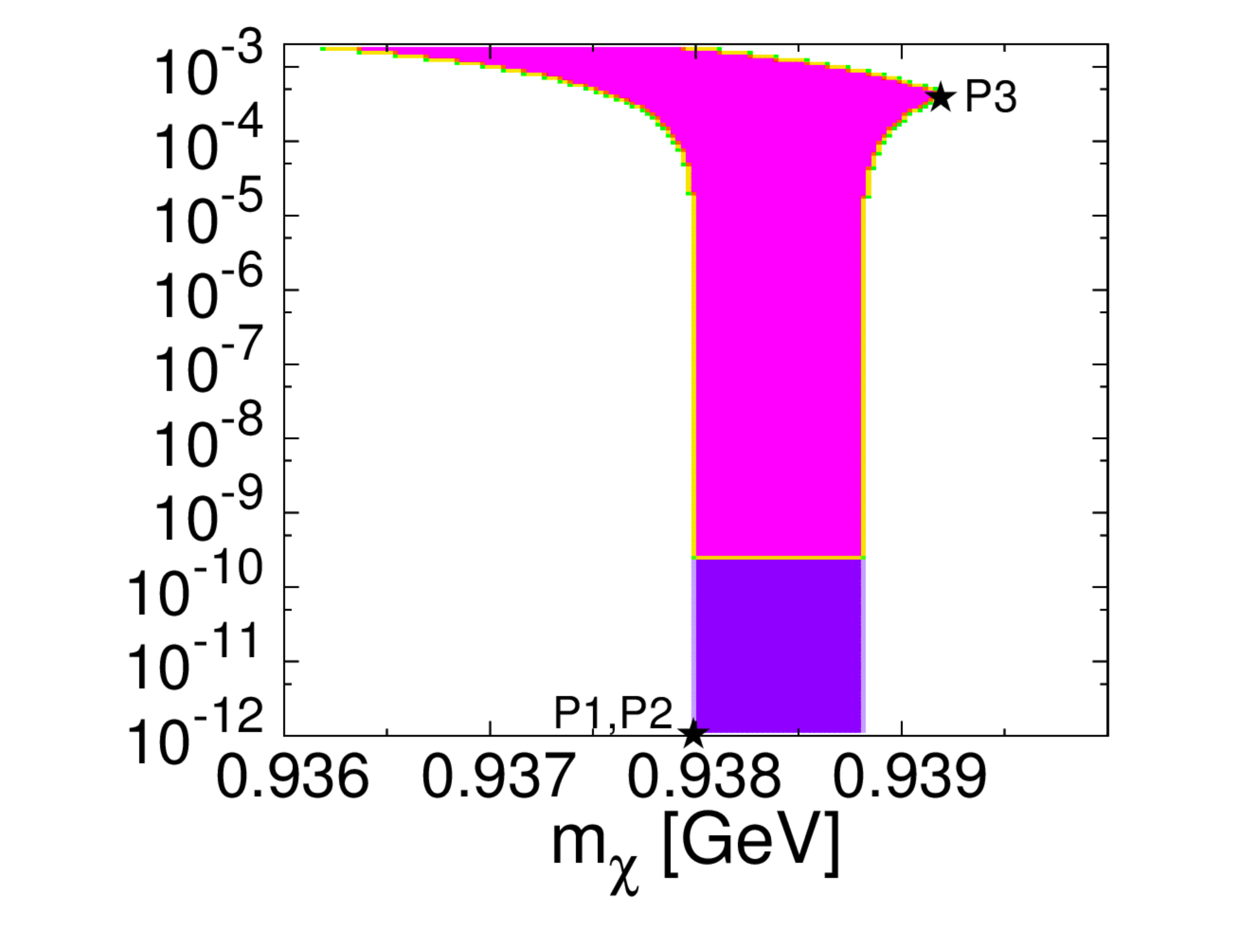}
\includegraphics[height=1.5in,angle=0]{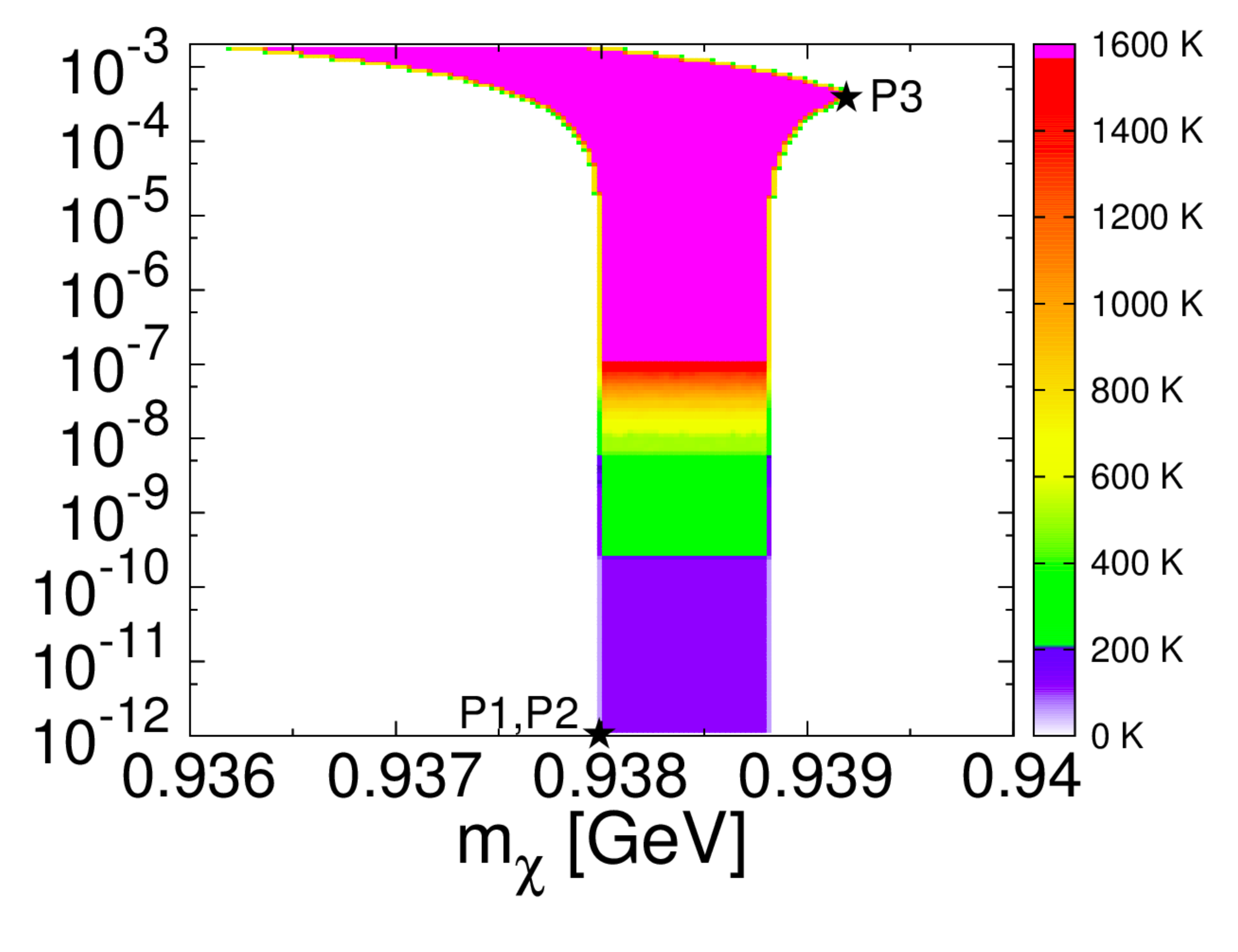}
\caption{\small \label{fig:fit_DM_modelII_case1}
 The minimum value of $T_{\rm obs}$ projected on the
$(m_\chi,m_\phi)$ plane when $\chi$ is DM.
The temperature scale is shown on the right panel. 
The stars mark the three benchmark points 
{\bf P1}, {\bf P2}, and {\bf P3}.
Left-panel: attractive DM self-energy $-\frac{n^2_\chi}{2z'^2}$.
Middle-panel: no DM self-energy.
Right-panel: repulsive DM self-energy $+\frac{n^2_\chi}{2z'^2}$.
}
\end{figure}

In this subsection, we consider the case in which $\chi$ is DM,
so there are no DM-neutron and DM-antiDM annihilation processes 
involved.
Figure~\ref{fig:fit_DM_modelII_case1}
shows the temperature increase in neutron stars older  than $10^{9}$~years in the parameter region of
Eq.~(\ref{eq:parameter}).
The panels from left to right correspond to 
attractive DM self-interaction,
no DM self-interaction, and repulsive DM self-interaction scenarios.
For each panel, the higher temperature region corresponds to a mixed phase of NS,
and the lower temperature region corresponds to the neutron phase.
A dramatic temperature change occurs at the boundary of these two phases.
For attractive DM self-interactions and no DM self-interactions, the boundary occurs for
 $m_\phi\simeq 0.2$~eV, which corresponds to $z\simeq 100$ MeV.
For repulsive DM self-interactions, the phase transition gradually occurs for
$10~{\rm eV}\lesssim  m_\phi \lesssim 100~{\rm eV}$, 
which corresponds to $5~{\rm GeV} \lesssim z \lesssim 50~{\rm GeV}$.

In the neutron phase, DM capture relies primarily on DM-neutron scattering.
We can see that the NS temperature is always below 200~K.
Because the DM-neutron cross section is too small 
to saturate the geometric limit,  the kinematic recoil energy of halo DM cannot heat up the NS.
In the mixed phase, there are substantial $\chi$ from neutron conversion inside the NS, 
and so, the DM self-capture kicks in and dramatically 
enhances the halo DM capture rate to the geometric limit.
This results in an observed NS temperature of 1580~K,
when the equilibrium condition 
$
L_\gamma|_{T_{\rm sur}=1660\,{\rm K}} = C_c|_{\rm geom}(\langle E_R \rangle) 
$ is satisfied.

\subsubsection{$\bar{\chi}$ is DM}

In this subsection, we assume $\bar{\chi}$ is the DM candidate.
Therefore, additional DM-neutron and DM-antiDM annihilation processes 
enhance the NS heating.

The DM-antiDM annihilation is through 
the $\chi \bar{\chi} \to \phi \phi$ process.
Whether or not $\chi \bar{\chi} \to \phi \phi$ enhances the NS temperature, depends on whether or not the decay products 
of $\phi$ can be absorbed by the NS.
If $\phi$ mixes with SM Higgs according to Ref.~\cite{Grinstein:2018ptl}, $\phi \to \gamma \gamma$ is the dominant channel,
so that NS heating can be further enhanced.
For scenarios in which $\phi$ decays 
into neutrinos or dark sector particles, DM-antiDM annihilation does not contribute 
to the heating process.
In the upper and lower rows of Fig.~\ref{fig:fit_DM_modelII_case23}, 
we separately show the two scenarios in which the final state particles are
absorbed or  not absorbed by the NS.

\begin{figure}[t!]
\centering
\includegraphics[height=1.40in,angle=0]{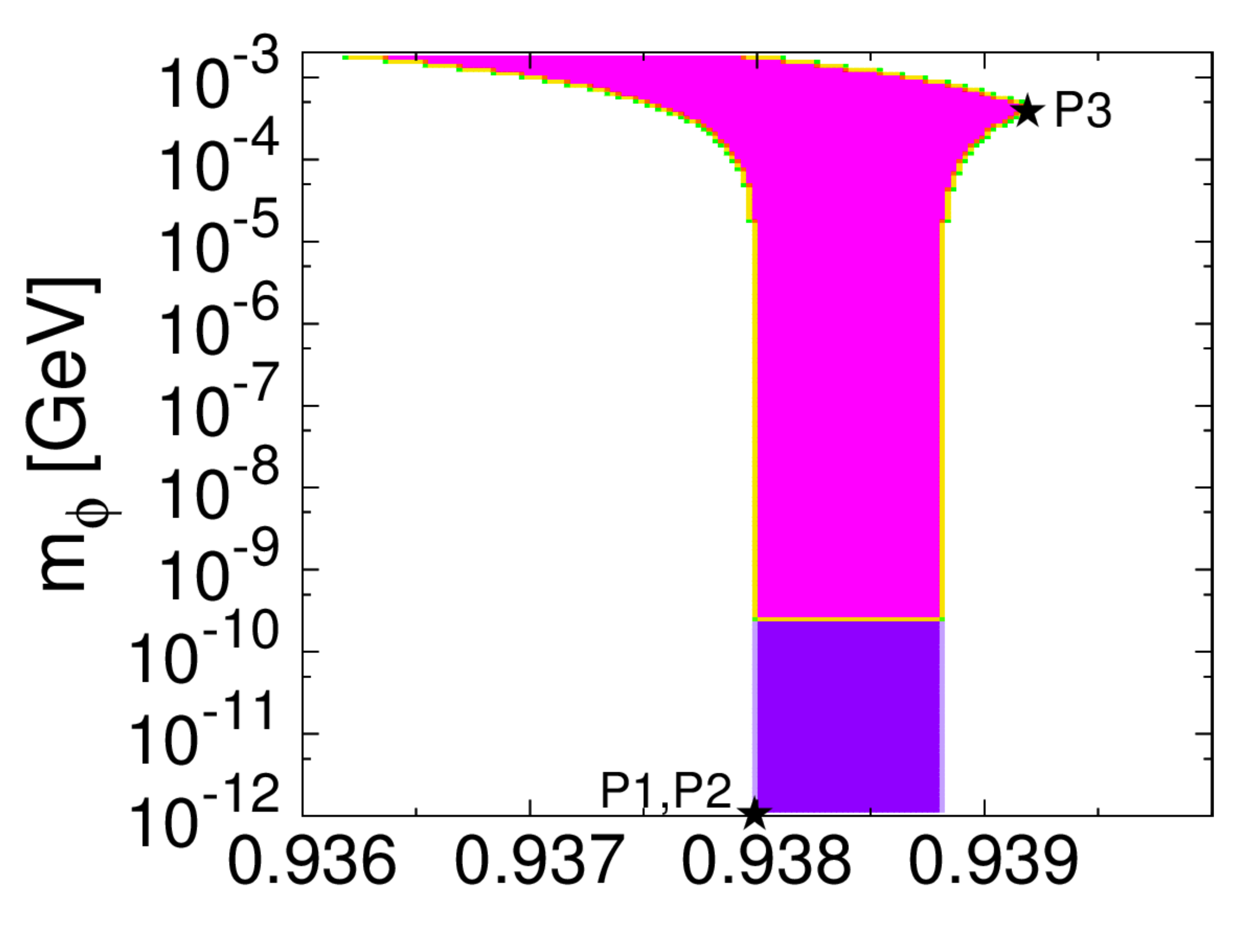}
%\hbox to 0.0in{}
\includegraphics[height=1.40in,angle=0]{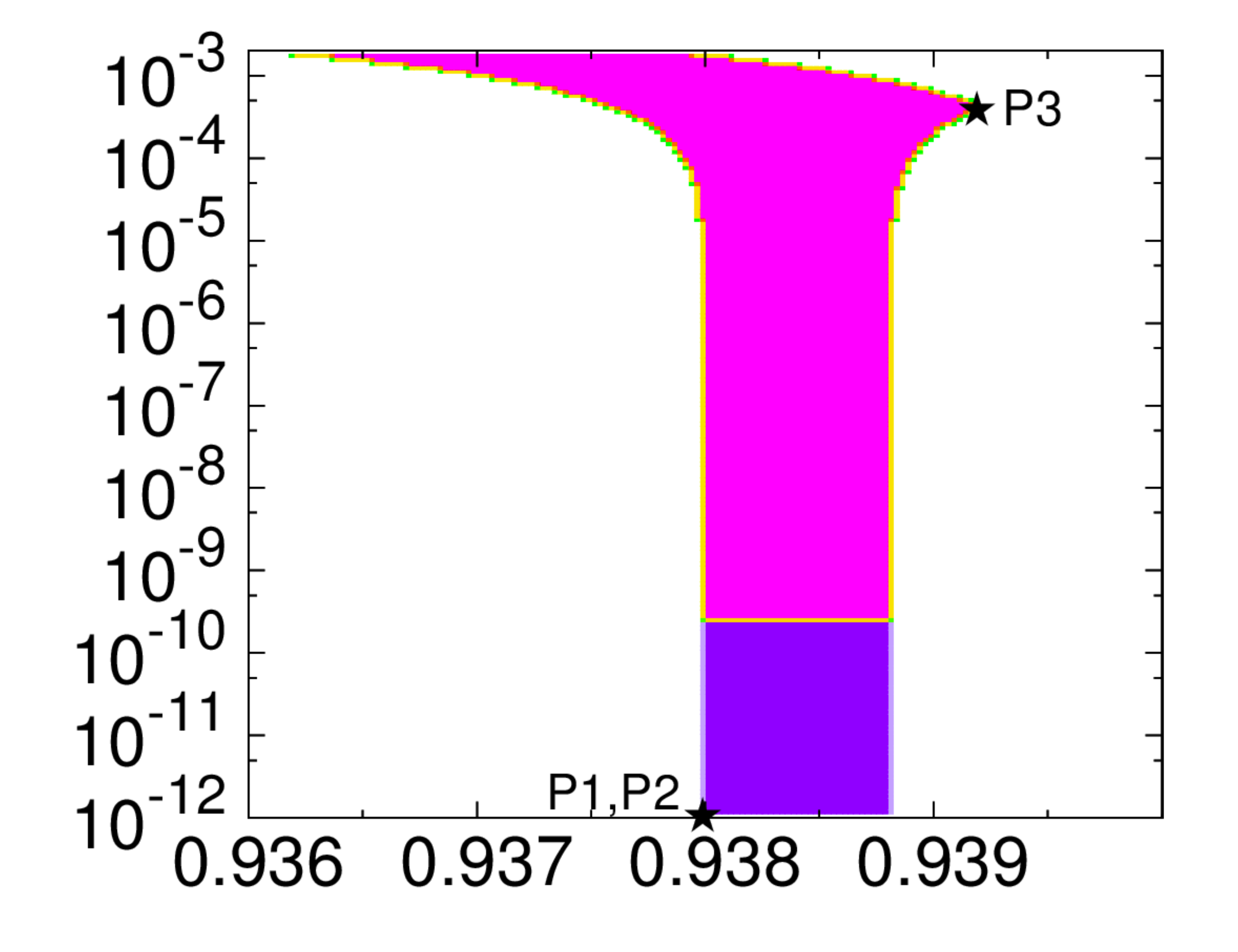}
%\hbox to 0.0in{}
\includegraphics[height=1.40in,angle=0]{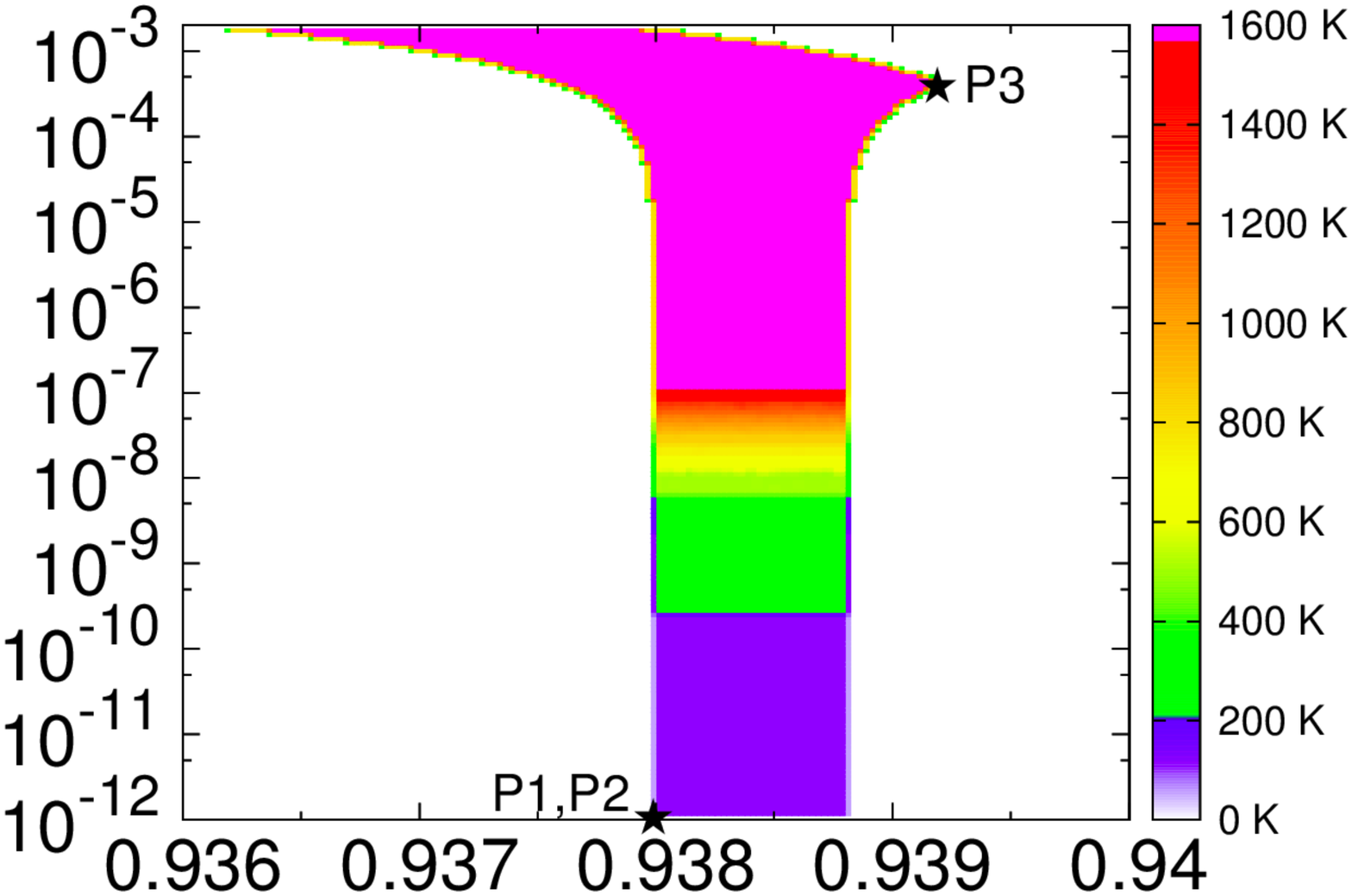}
%\hbox to 0.0in{}
\includegraphics[height=1.55in,angle=0]{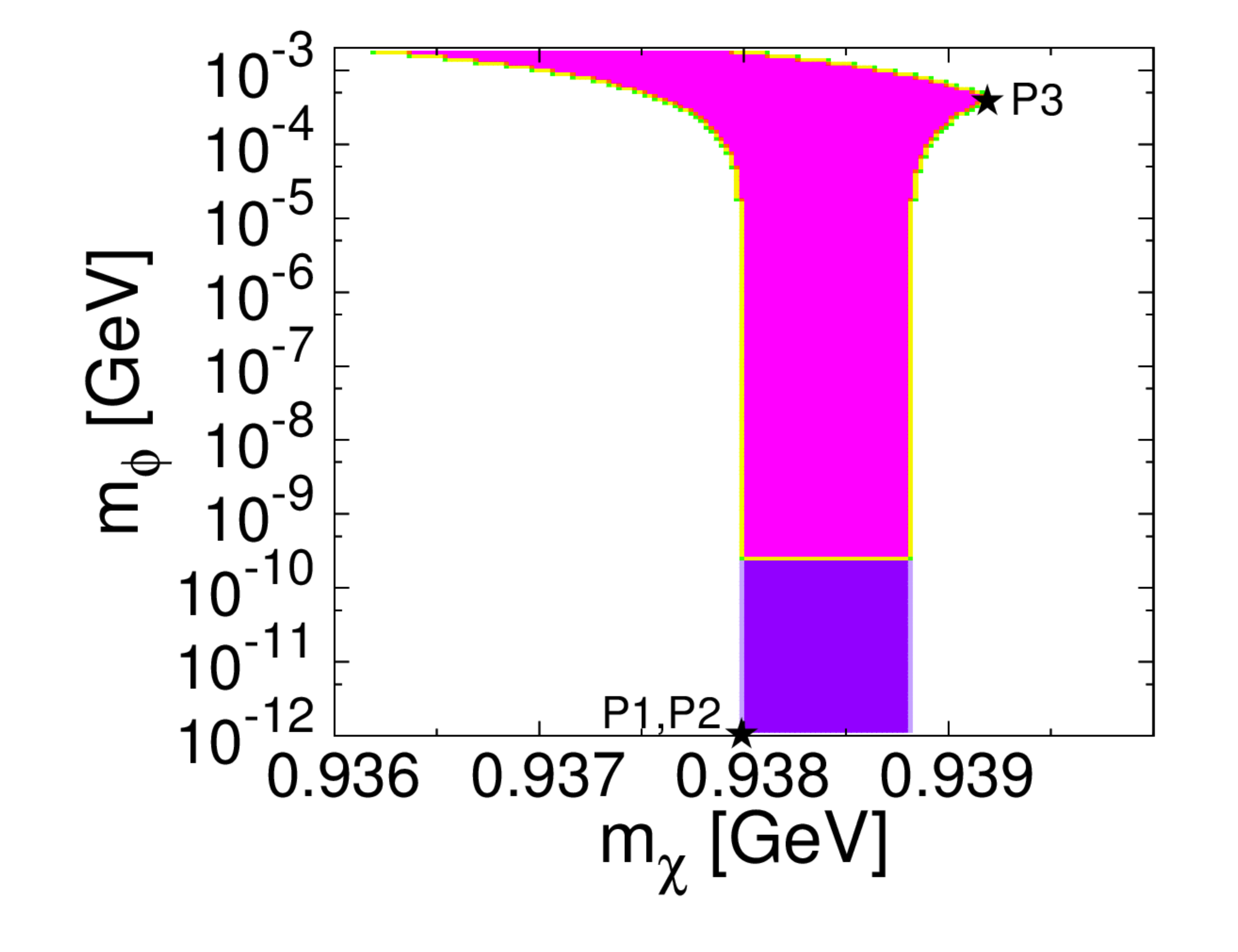}
%\hbox to 0.0in{}
\includegraphics[height=1.55in,angle=0]{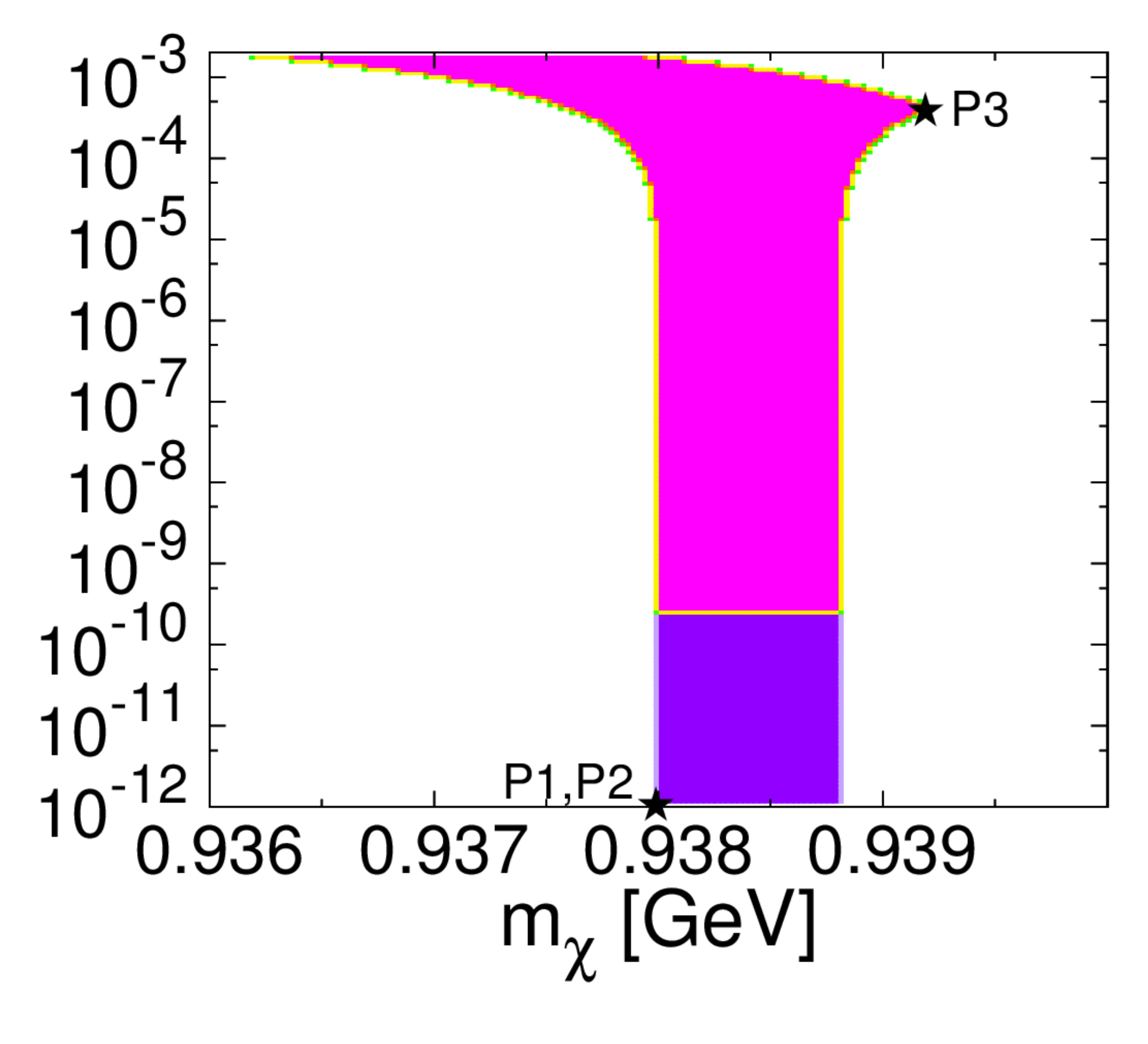}
\hbox to 0.0in{}
\includegraphics[height=1.55in,angle=0]{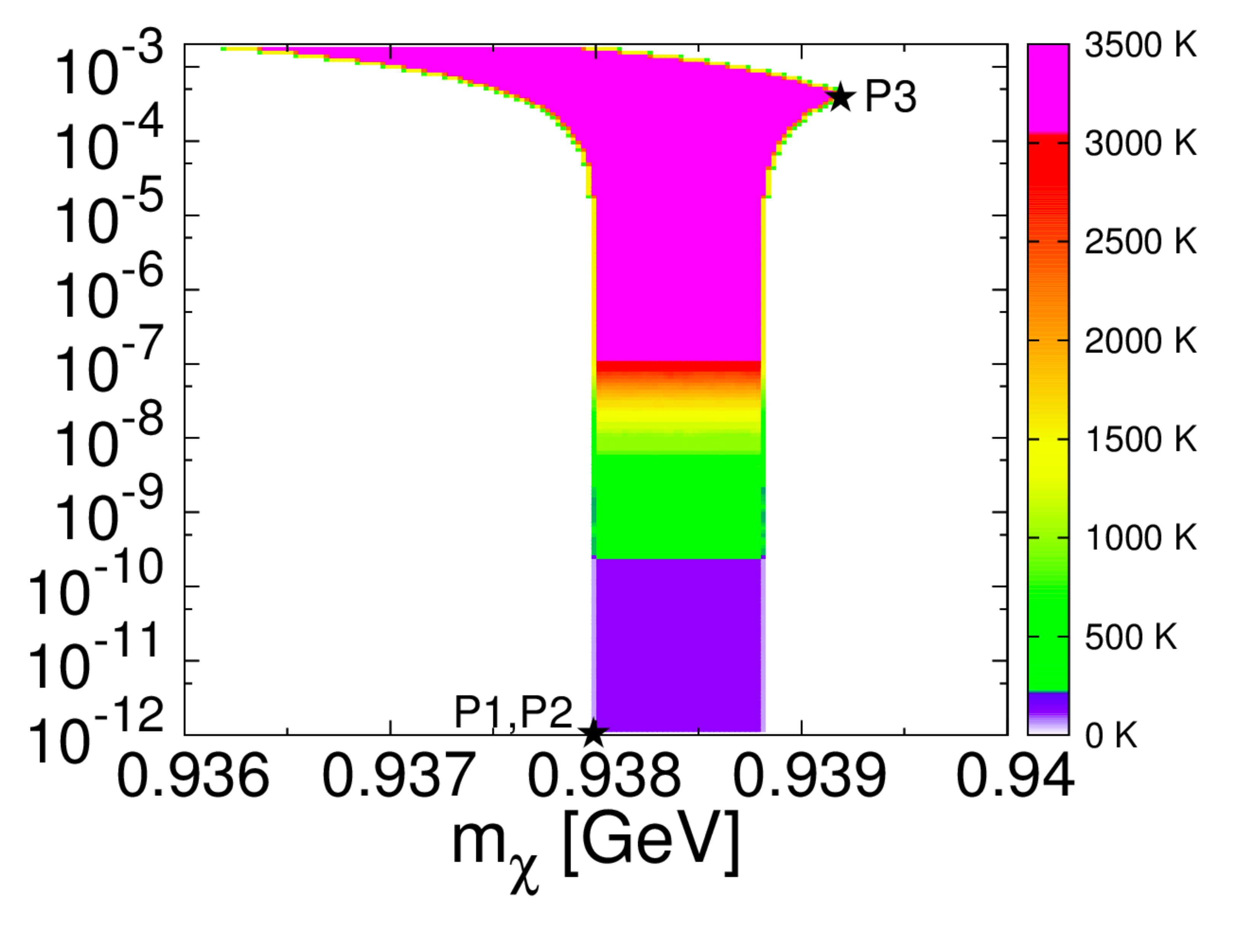}
\hbox to 0.03in{}
\caption{\small \label{fig:fit_DM_modelII_case23}
Same as Fig.~\ref{fig:fit_DM_modelII_case1} 
but $\bar{\chi}$ is DM.
Upper-row: $\phi$ decay final states cannot be absorbed by the NS.
Lower-row: $\phi$ decay final states are absorbed by the NS.
}
\end{figure}

In Fig.~\ref{fig:fit_DM_modelII_case23}, for each panel,
there are higher and lower temperature regions respectively 
corresponding to the mixed and neutron phases. 
The upper row of Fig.~\ref{fig:fit_DM_modelII_case23},
which shows the neutron phase, has an additional DM-neutron annihilation process (compared to the $\chi$ DM case)
to heat up the NS.
However, its contribution is insignificant and the observed temperature is below 200~K.
In the mixed phase, again, the substantial component of $\chi$ in the NS 
and large DM-antiDM scattering help the capture rate 
to reach the geometric limit,
but the additional DM-$\chi$ annihilation cannot heat up the NS, 
because the annihilation final states cannot been absorbed.
The result is that kinetic heating raises the NS temperature to 1580~K.

In the mixed phase, the DM-antiDM annihilation process enhances the NS observed temperature up to 3100~K 
corresponding to a surface temperature of 3440~K; see the lower panel of Fig.~\ref{fig:fit_DM_modelII_case23}.
This occurs when the equilibrium condition 
$
L_\gamma|_{T_{\rm sur}=3440\,{\rm K}} = C_c|_{\rm geom}(\langle E_R \rangle+2m_\chi) 
$ is satisfied.
But in the neutron phase, the temperature is lower than 200~K
because there is no $\chi$ component from neutron conversion 
to annihilate with DM $\bar{\chi}$.

\bigskip

\section{Quark vector current portal dark matter}
\label{sec:vector_portal}

We consider Dirac DM with mass around a GeV that couples to quarks 
through a vector current interaction.
It is difficult for current DM direct detection experiments to probe this scenario
because the recoil energy is much lower than the typical detector threshold. 
However,  the leading DM annihilation final state is $\pi^+\pi^-$, 
 which produces MeV photons that can be observed by near future instruments that will fill in the
``MeV-gap" in the cosmic photon spectrum~\cite{Berger:2019aox,Kumar:2018heq}.
Through the quark vector current, we also expect substantial DM-neutron 
scattering that will enable a NS to capture halo DM, which in turn will heat the NS.

\bigskip

\subsection{DM-nucleon scattering cross section}

Consider a Dirac fermion DM particle $\chi$ that couples to quarks through a vector-vector current,
\begin{eqnarray}
\label{eq:0}
\mathcal{L}_{int}=\sum_{q=u,d,s} \frac{\alpha_q}{\Lambda^2}\bar{\chi}\gamma^\mu \chi \bar{q}\gamma_\mu q\,,
\end{eqnarray}
where $\alpha_q$ are the coupling strengths and $\Lambda$ is a cutoff scale.
To describe DM capture by a NS, the DM-neutron scattering cross section 
should be calculated in the relativistic limit, 
since the DM particles are accelerated close to the speed of light.
The DM-neutron and DM-proton cross sections are given by
~\cite{Bell:2018pkk,Feng:2016ijc}
\begin{eqnarray}
\frac{d \sigma_{\chi n,p}(s,t)}{d\cos\theta_{\rm cm}}=
\frac{c_{\chi n,p}}{\Lambda^4} 
\frac{2(\bar{\mu}^2+1)^2m^4_\chi -4(\bar{\mu}^2+1)\bar{\mu}^2 s m^2_\chi+\bar{\mu}^4 (2s^2+2st+t^2)}{16\pi \bar{\mu}^4 s}
|F_n(E_R)|^2\,, \nonumber \\
\end{eqnarray}
where $\theta_{\rm cm}$ is the scattering angle in the center mass frame and 
$\bar{\mu}\equiv m_\chi/m_n\simeq m_\chi/m_p$.
Here, $c_{\chi p,n}=(\alpha_u B^{p,n}_u+\alpha_d B^{p,n}_d)^2$,
with the integrated nuclear form-factors, $B^p_u=B^n_d=2$ 
and $B^n_u=B^p_d=1$.
The nucleon form factor is 
$
|F_n(E_R)|^2=\exp[-E_R/(0.114~{\rm GeV})]
$
\cite{Feng:2016ijc}, where $E_R$  
is the recoil energy in the initial $n$ or $p$ rest frame.
For DM capture by a NS,
in the initial nucleon rest frame,
the energy of DM due to gravitational acceleration is $m_\chi/\sqrt{1-\omega^2}\simeq m_\chi/\sqrt{\bar{B}}$, where we have neglected
the thermal motion of the DM.
The expressions for the other kinematic variables are
\begin{eqnarray}
&& s = m^2_\chi+m^2_n+2m_\chi m_n/\sqrt{\bar{B}}\,, \nonumber \\
&& t =-2|\overrightarrow{p_0}|^2(1-\cos \theta_{\rm cm})\,, \nonumber \\
&& E_R = \frac{|\overrightarrow{p_0}|^2}{m_{n,p}}(1-\cos \theta_{\rm cm})\,, \nonumber \\
&& |\overrightarrow{p_0}|^2 = \frac{(1-\bar{B})m_\chi m_n \bar{\mu}}{\bar{B}+2\sqrt{\bar{B}}\bar{\mu}+\bar{B}\bar{\mu}^2}
\end{eqnarray}
where 
$
|\overrightarrow{p_0}|=\frac{\sqrt{s}}{2}\lambda^{1/2}(1,m^2_\chi/s,m^2_n/s)
$
%$\bar{B}\equiv 1-2GM/(c^2 R)$,
and $\lambda(x,y,z)\equiv x^2+y^2+z^2-2xy-2xz-2yz$.

An example of the DM-neutron scattering cross section 
for DM capture by a NS is provided in Fig.~\ref{fig:ann_KpKn}.
By choosing couplings strengths $\alpha_q=\mathcal{O}(10^{-4})$ 
and $\Lambda=100$~GeV,
the DM-neutron cross section is larger than the critical cross section. 
Therefore, we expect
the corresponding DM capture rate to reach the geometric limit
and an old NS temperature can be heated up to 1500~K.
The sensitivity provided by NS heating is significantly greater than 
that from future observations 
of MeV cosmic photons by e-ASTROGAM, AMEGO and APT, 
which are sensitive to $\alpha_q/\Lambda \sim \mathcal{O}(1)/100$~GeV~\cite{Berger:2019aox}.

\begin{figure}[t!]
\centering
\includegraphics[height=2.9in,angle=270]{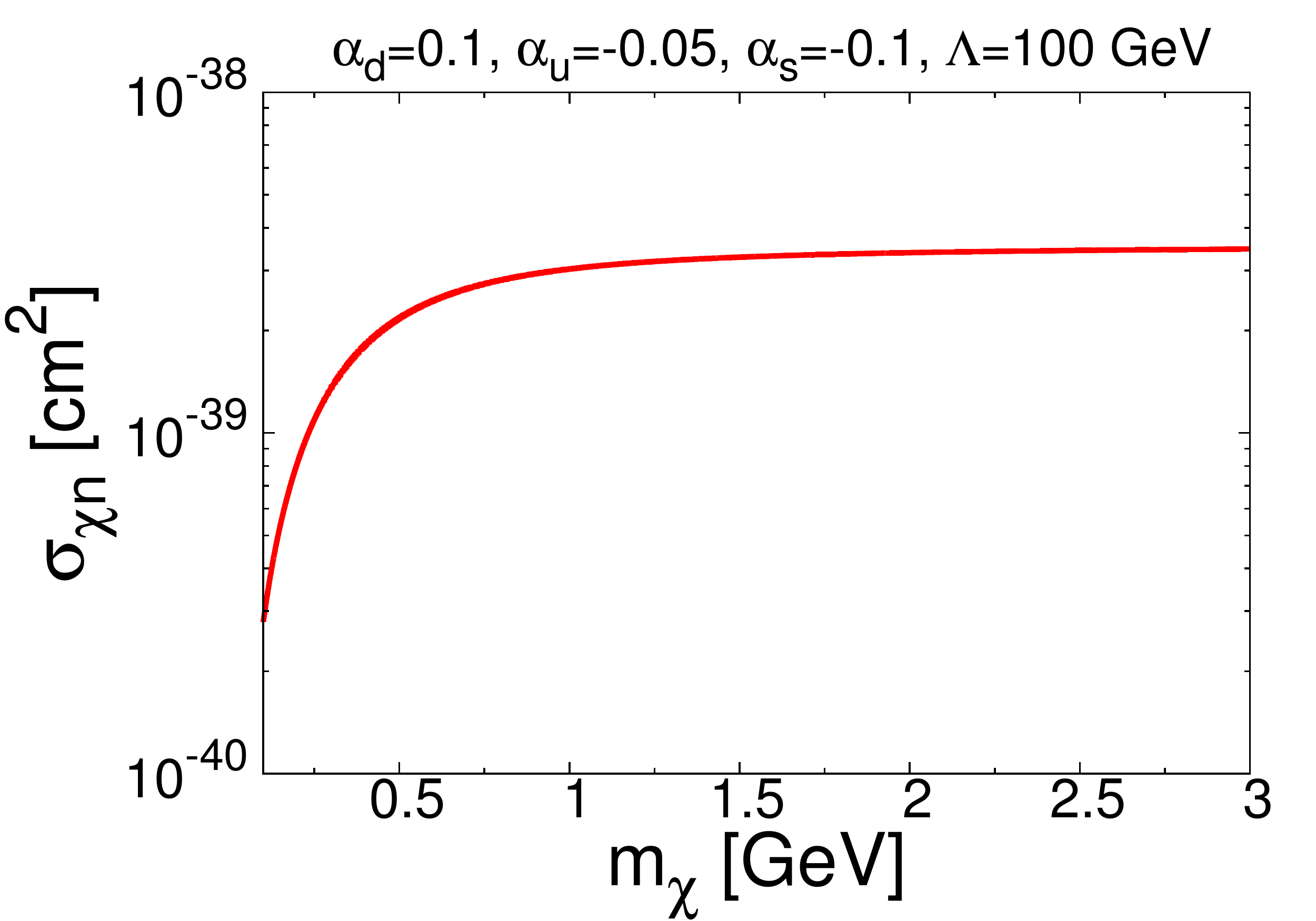}
\includegraphics[height=2.9in,angle=270]{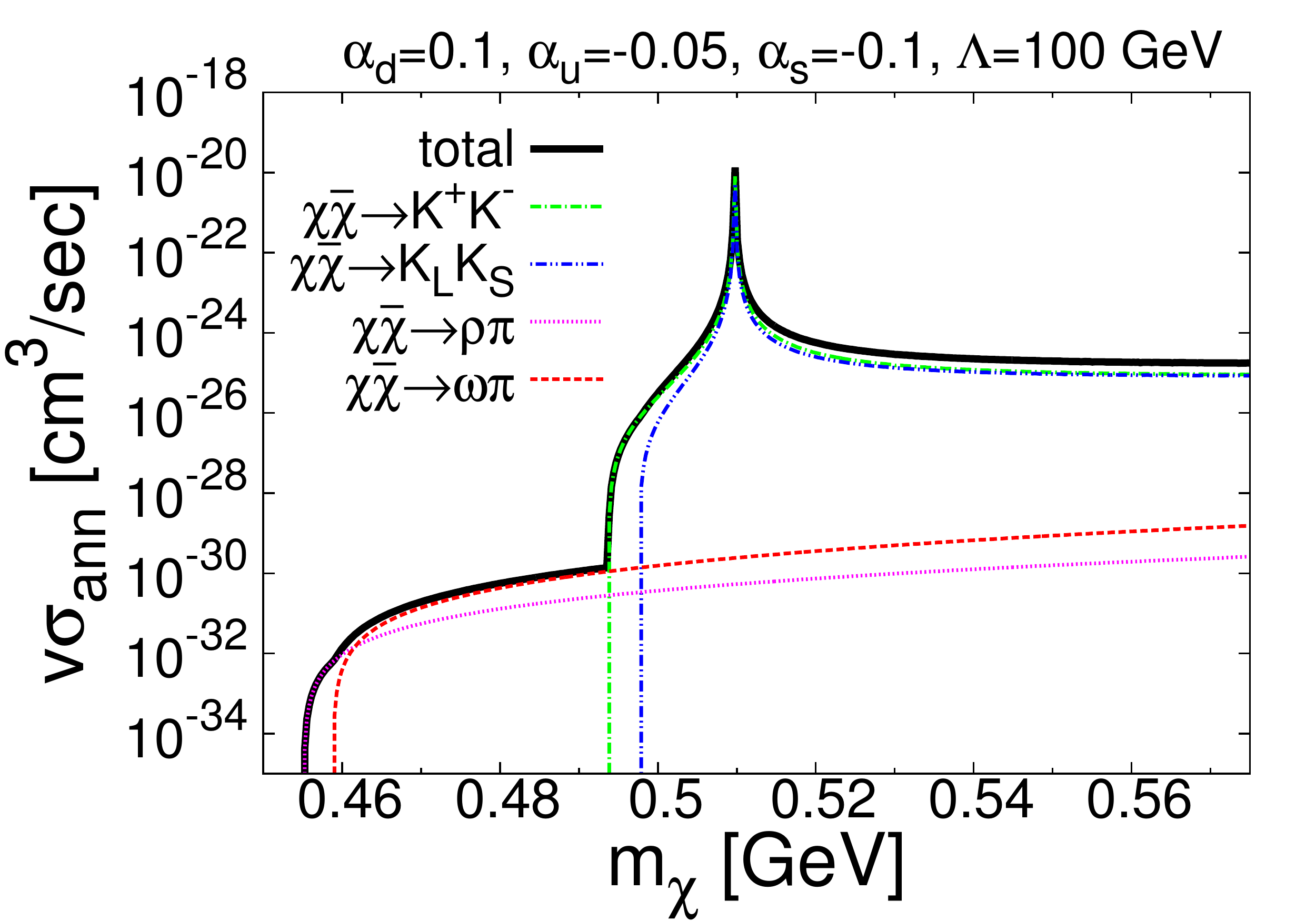}
\caption{\small \label{fig:ann_KpKn}
Left-panel: The DM-neutron scattering cross section in the relativistic limit
for DM capture by a NS with $M=1.44 M_\odot$ and $R=10.6$ km.
Right-panel: The $\chi \bar{\chi}$ annihilation cross sections
for $\sqrt{s}\leq 1.15$~GeV including interference effects.
}
\end{figure}

\bigskip

\subsection{Chiral Lagrangian and DM annihilation}

We now calculate NS heating due to DM-antiDM annihilation.
At the GeV scale, DM-quark vector current interactions 
can be described by Chiral perturbation theory,
such that the DM annihilate into pseudoscalar or vector mesons.
We focus on $\sqrt{s}\lesssim 1.15$~GeV, so that we only need to include the
$\chi \bar{\chi} \to K^+K^-,K_LK_S,\rho \pi,\omega \pi$ channels.

The Feynman rules for GeV DM couplings to low-energy QCD pseudoscalar meson 
and vector meson can be found in appendix B of Ref.~\cite{Berger:2019aox}.
Then the vector meson propagator
$<0|T(\rho_{\mu\nu},\rho_{\alpha\beta})|0>$ is~\cite{Ecker:1988te}
\begin{eqnarray}
\label{eq:1}
\frac{g_{\mu \alpha} g_{\nu\beta}(m^2_\rho -k^2)
+{g_{\mu\alpha}k_\nu k_\beta}
-{g_{\mu\beta}k_\nu k_\alpha}
-g_{\nu \alpha} g_{\mu\beta}(m^2_\rho -k^2)
-{g_{\nu\alpha}k_\mu k_\beta}
+{g_{\nu\beta}k_\mu k_\alpha}
}
{( m^2_\rho)(m^2_\rho-k^2-i\varepsilon)}\,,
\end{eqnarray}
and the polarization of $\rho_{\mu\nu}$ is
$\left[ k_\mu \epsilon_\nu(k)-k_\nu \epsilon_\mu(k) \right]/m_\rho$.
The polarization sum between $\rho_{\mu\nu}$ and $\rho_{\mu'\nu'}$ is given by
$
\left[
k_\mu k_{\nu'}g_{\nu \mu'}
+k_\nu k_{\mu'}g_{\mu \nu'}
-\left( k_\mu k_{\mu'}g_{\nu \nu'}+k_\nu k_{\nu'}g_{\mu \mu'} \right) 
\right]/m^2_\rho
$.
Using the $\rho$ propagator in Eq.~(\ref{eq:1}) and the $\chi\bar{\chi}\rho$, $K^+K^-\rho$ vertices from Appendix B of 
Ref.~\cite{Berger:2019aox}, the amplitude squared for 
$\chi(p) \bar{\chi}(p')\to \rho \to K^+(k)K^-(k')$ is
\begin{eqnarray}
\frac{1}{4}\sum |M|^2 &=& (\alpha_d-\alpha_u)^2 
   \left\lbrace \frac{4f^2_V h_p}{2 \Lambda^2 F^2}  \right\rbrace^2 
   \frac{1}{(s-m^2_\rho)^2 +m^2_\rho \Gamma^2_\rho} \nonumber \\
   &\times & 2 \left[ s^2-4 s\, m^2_K-(u-t)^2 \right]
                \left[ (u+t)-2(m^2_\chi+m^2_K) \right]^2\,,
\end{eqnarray}
where $s\equiv(p+p')^2=(k+k')^2$, $t\equiv(p-k')^2=(k-p')^2$, 
$u\equiv(p-k)^2=(k'-p')^2$, 
and the values for $f_V$, $h_p$, and $F$ can be found 
in Ref.~\cite{Berger:2019aox}
In terms of the Mandelstam variables, 
\begin{eqnarray}
u-t &=& -4|\overrightarrow{p}||\overrightarrow{k}|\cos\theta\,, \nonumber \\ 
u+t &=& -2(|\overrightarrow{p}|^2+|\overrightarrow{k}|^2)\,, \nonumber
\end{eqnarray}
where $\theta$ is the angle between $\overrightarrow{p}$ and $\overrightarrow{k}$,
and $|\overrightarrow{p}|=\frac{\sqrt{s}}{2} \sqrt{1-\frac{4m^2_\chi}{s}}$,  $|\overrightarrow{k}|=\frac{\sqrt{s}}{2} \sqrt{1-\frac{4m^2_K}{s}}$.
In the threshold limit, $s\rightarrow 4m^2_\chi\Rightarrow $ $u-t=0$ and $u+t=-2 (m^2_\chi-m^2_K)$.
Then the amplitude squared can be simplified to
\begin{eqnarray}
\frac{1}{4}\sum |M|^2 &=& (\alpha_d-\alpha_u)^2 
   \left\lbrace \frac{4f^2_V h_p}{2 \Lambda^2 F^2}  \right\rbrace^2 
   \frac{1}{(4m^2_\chi -m^2_\rho)^2 +m^2_\rho \Gamma^2_\rho} \nonumber \\
   &\times & 512\, m^8_\chi \left( 1-\frac{m^2_K}{m^2_\chi} \right)\,.
\end{eqnarray}
The total and partial $\bar{\chi}\chi$ annihilation cross sections 
are shown in Fig.~\ref{fig:ann_KpKn} including interference effects.
For $\sqrt{s}> 1.15$ GeV, 
other channels are kinematically viable,
like a glueball with neutral pions.
Because the calculation of glueball emission is beyond the scope of this work,
we only consider the DM annihilation cross section for $\sqrt{s}\lsim 1.15$~GeV.
Moreover, as long as the DM annihilation rate is large enough to 
maintain equilibrium 
between the DM capture rate and depletion rates, 
including the new channels 
do not further increase the temperature of the NS. 
Without including the DM annihilation channels above $\sqrt{s}=$1.15~GeV, 
we still obtain a conservative estimate of NS heating 
for DM masses above 0.575~GeV.

\subsection{Results}
\label{sec:vector_results}

\begin{figure}[t!]
\centering
\includegraphics[height=3.2in,angle=0]{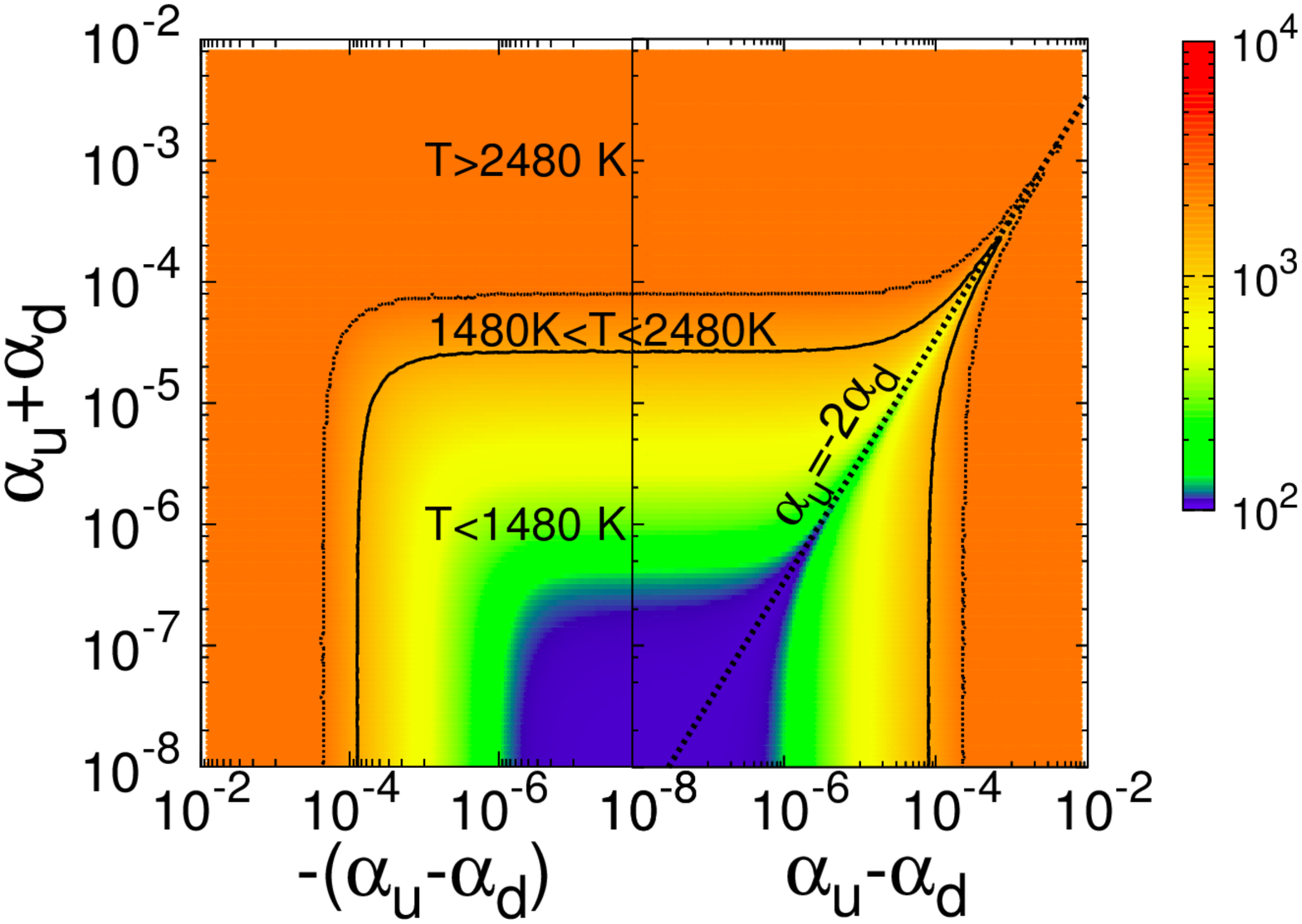}
\caption{\small \label{fig:auad}
$T_{\rm obs}$ (in K) in the vector portal DM framework by varying $\alpha_u$ and $\alpha_d$.
We fix $\alpha_s=0$, $\Lambda=100$~GeV and $m_\chi=1$~GeV.
%The curves show $T_{\rm obs}=1480$ K and 2480 K of NS, and the color represents other values of $T_{\rm obs}$.
%The dashed line indicates $\alpha_u=-2 \alpha_d$.
}
\end{figure}

In Fig.~\ref{fig:auad}, we shown the observed temperature of the NS due to the vector-vector current couplings to quarks in
Eq.~(\ref{eq:0}). For $\alpha_u$ or $\alpha_d$ larger than ${O}(10^{-4})$, 
DM capture heats up the NS to more than 1480~K, 
which is shown by the black curve.
However, for $\alpha_u=-2 \alpha_d$ the DM-neutron scattering cross section vanishes, and the NS does not get heated. 
This feature is indicated by the dashed line in Fig.~\ref{fig:auad}.

In Fig.~\ref{fig:as_auad}, we vary $\alpha_s$ and $\alpha_u+\alpha_d$, 
and fix $\alpha_u=\alpha_d$, $\Lambda=100$~GeV and $m_\chi=1$~GeV.
Clearly, $T_{\rm obs}$ 
is insensitive to the parameter $\alpha_s$, 
which modifies $v\sigma^{\rm ann}$, but not $\sigma^{\rm elastic}_{\chi n}$;
$\alpha_s$ only affects $\mathcal{O}(100)$~K temperatures.
We may understand the features of Fig.~\ref{fig:as_auad} as follows.
First, focus on the region of $T_{\rm obs}$ above 1000~K, 
where  $\sigma^{\rm elastic}_{\chi n}$ is close to $\sigma_{\rm crit}\simeq 2\times 10^{-45}\,{\rm cm^2}$ 
and the DM capture rate $C_c$ reaches the geometric limit.
This corresponds 
to $\alpha_u\simeq \alpha_d\simeq 4\times 10^{-5}$ 
which gives a DM annihilation cross section, $v\sigma^{\rm ann}\simeq \mathcal{O}(10^{-33})\,{\rm cm^3/s}$, 
which is six orders of magnitude larger than $v\sigma^{\rm ann}|_{\rm sat}\simeq \mathcal{O}(10^{-39})\,{\rm cm^3/s}$.
$\alpha_s$  only alters $v\sigma^{\rm ann}$ within a similar magnitude, but 
cannot suppress it down to $v\sigma^{\rm ann}|_{\rm sat}$.
Thus, for $T_{\rm obs}$ around 1000~K, $T_{\rm obs}$ is insensitive to $\alpha_s$.

\begin{figure}[t!]
\centering
\includegraphics[height=3.2in,angle=0]{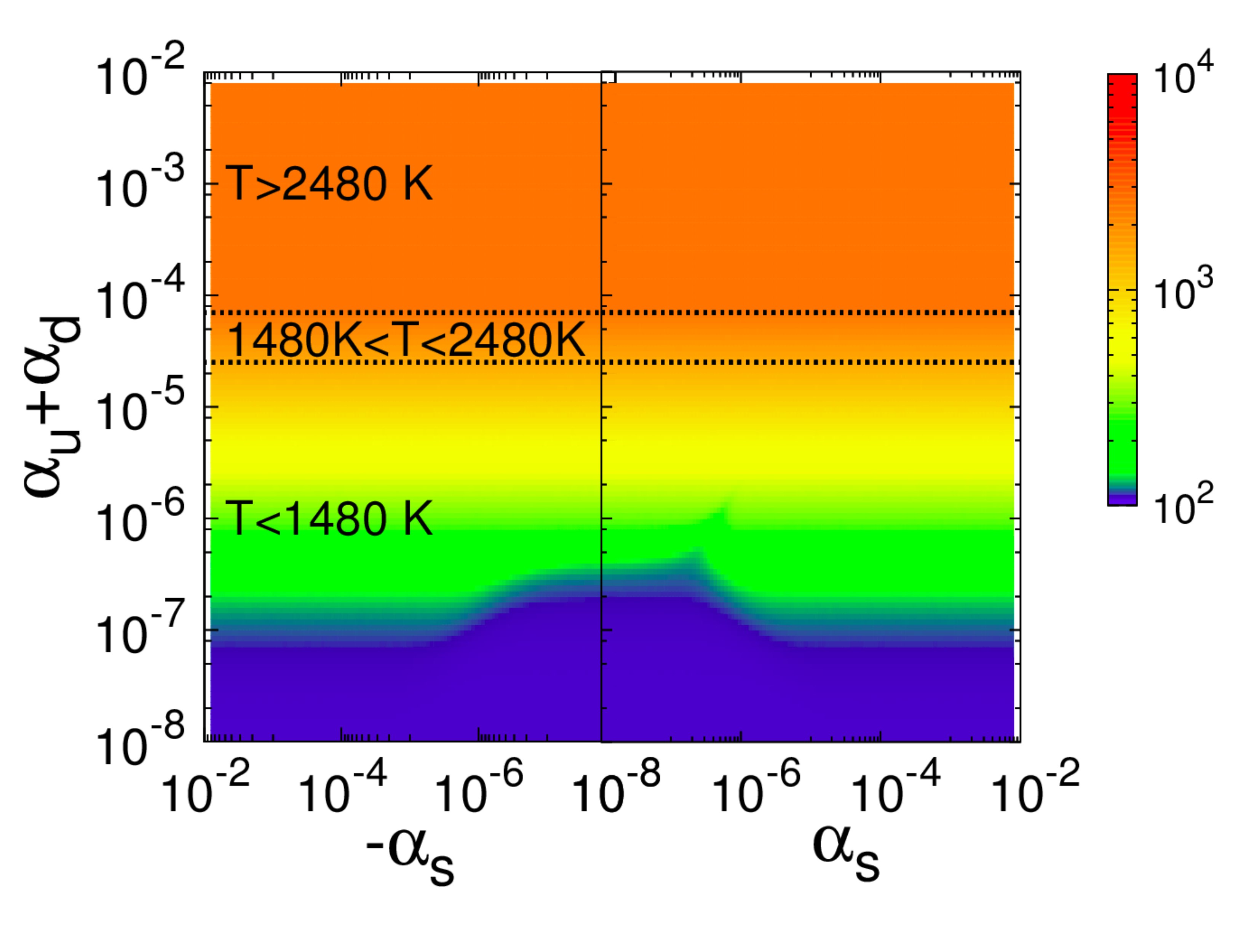}
\caption{\small \label{fig:as_auad}
$T_{\rm obs}$ (in K) in the vector portal DM framework by varying $\alpha_s$ and $\alpha_u+\alpha_d$, 
while fixing $\alpha_u=\alpha_d$, $\Lambda=100$~GeV and $m_\chi=1$~GeV.
%The dashed lines show $T_{\rm obs}=1480$ K and 2480 K of NS, and the color represents other values of $T_{\rm obs}$.
}
\end{figure}

However, the situation is different when the final $T_{\rm obs}$ is of 
$\mathcal{O}(100)$~K,
which corresponds to much smaller values of $C_c$ and $\sigma^{\rm elastic}_{\chi n}$.
Take $C_c= 10^{-4}\times C_c|_{\rm geom}$  as an example.
This corresponds to $\sigma^{\rm elastic}_{\chi n}=2\times 10^{-49}\,{\rm cm^2}$ 
and $\alpha_u\simeq \alpha_d\simeq 4\times 10^{-7}$, which gives
$v\sigma^{\rm ann}\simeq \mathcal{O}(10^{-37})$, 
which is much smaller than the saturating annihilation cross section, $v\sigma^{\rm ann}|_{\rm sat}\simeq \mathcal{O}(10^{-35})$.
This means that increasing $v\sigma^{\rm ann}$ by varying $\alpha_s$  enhances
the final NS temperature $T_{\rm obs}$.
This behavior at $\mathcal{O}(100)$~K is evident from the dark blue region in Fig.~\ref{fig:as_auad}.
For  $\alpha_u+\alpha_d=2\times 10^{-7}$, increasing $|\alpha_s|$ from $10^{-8}$ to $10^{-5}$ raises $T_{\rm obs}$, which plateaus for
 $|\alpha_s|>10^{-5}$.
The little spike at $\alpha_s\simeq 3\times 10^{-7}$ 
is due to destructive interference
between the DM annihilation channels.

\bigskip

\section{Summary}
\label{sec:summary}

We have investigated NS heating by the capture of GeV-mass DM.
We discussed the generic scenario that the NS could be in a mixed phase composed of both neutrons and a 
substantial  population of DM from neutron conversion.
In this case, the geometric limit of the DM capture rate can be saturated through 
DM self-interactions without DM-neutron interactions.

A NS can be in a mixed phase in the neutron dark decay model 
(that explains the neutron lifetime anomaly), 
because neutrons are able to convert to DM.
We demonstrated that a NS in mixed phase can be stable and its mass can be as heavy as 2$M_\odot$ 
by solving the equation of state and Tolman-Oppenheimer-Volkoff equations.

To illustrate the effect of DM capture on NS heating,
we chose the above mentioned neutron dark decay model 
and the quark vector current portal framework.
For the neutron dark decay model, since the DM self-scattering cross section 
is crucial to estimate the DM capture rate,
we calculated the tree-level and one-loop box diagram contributions.
In the mixed phase of a NS, DM self-scattering 
can enhance the DM capture rate 
up to the geometric limit without DM-neutron interactions.
We find that for $m_\phi\gtrsim 100$~eV, 
the sensitivity of near future infrared instruments is greater
than afforded by multi-pion signatures 
at Super-Kamiokande, Hyper-Kamionkande, and DUNE.

For quark vector portal DM, since the NS is in the neutron phase, 
 halo DM is captured only via DM-neutron interactions.
We find that the capture rate
is close to the geometric limit for $\alpha_{u,d}\gtrsim \mathcal{O}(10^{-4})$, in which case the NS is heated to $\sim1500$~K.
This is four orders of magnitude more sensitive
than the detection of MeV cosmic gamma rays by e-ASTROGAM, AMEGO and APT,
which are sensitive to $\alpha_{u,d}\simeq \mathcal{O}(1)$~\cite{Berger:2019aox}.
We also find that  NS heating is not sensitive to $\alpha_s$,
unless future telescopes can observe NS temperatures of around 100~K.

A NS that is heated to 1480~K produces a
photon spectrum that is peaked at about 1-2~$\mu$m and has a spectral 
flux density of $\simeq 0.5$~nJy if the NS is at a distance of 10~pc from Earth. 
This is near the optimal sensitivity of the upcoming infrared telescopes, 
JWST, Thirty Meter Telescope, and European Extremely Large Telescope~\cite{Baryakhtar:2017dbj}.
JWST is closest to completion, and is
expected to reach $\mathcal{O}(10)$ signal-to-noise 
for $\mathcal{O}(10)$~nJy in a typical integration time of $10^4$ seconds~\cite{JWST}.
A 2480~K NS at 10~pc (50~pc) can be detected by JWST in 2000 seconds ($\mathcal{O}(10^6)$ seconds).

\section*{Acknowledgements}  
We thank J.~Kumar and X.~Tata for discussions. W.-Y.K. and P.-Y.T. thank the National Center of Theoretical Sciences, 
Taiwan, for its hospitality.  D.M. thanks the Aspen Center for
Physics (which is supported by U.S. NSF Grant No. PHY-1607611) for its hospitality while this work was in progress. D.M. is supported in
part by the U.S. DOE under Grant No. de-sc0010504.

%\newpage
%%%%%%%%%%%%%%%%%%%%%%%%%%%%%%%
%%%%%%%%%%% Appendix %%%%%%%%%%%
%%%%%%%%%%%%%%%%%%%%%%%%%%%%%%%
\appendix

%%%%%%%%%%%%%%%%%%%%%-------------------

\end{document}